\documentclass[aps,prb,reprint]{revtex4-1}

\usepackage{bm,hyperref}
\usepackage{amsmath}
\usepackage{graphicx}
\usepackage{soul}
\usepackage{amsfonts}
\usepackage{amssymb}
\usepackage{dcolumn}
\usepackage{amsmath}

\usepackage[T1]{fontenc}
\usepackage[latin9]{inputenc}

\usepackage{color}
\usepackage{float}
\usepackage{amsmath}
\usepackage{amssymb}
\usepackage{graphicx}
\usepackage{wasysym}
\usepackage{esint}

\begin{document}

\title{Theory of optical forces on small particles by multiple plane waves}


\author{Ehsan Mobini{$^{*,1}$}, Aso Rahimzadegan{$^{2}$}, Carsten Rockstuhl{$^{2,3}$}, and Rasoul Alaee{$^{*,4}$}}
\address{
{$^{1}$}Department of Physics, University of Ottawa, Ottawa, Canada\\
{$^{2}$}Institute of Theoretical Solid State Physics, Karlsruhe Institute of Technology, Karlsruhe, Germany\\
{$^{3}$}Institute of Nanotechnology, Karlsruhe Institute of Technology, Karlsruhe, Germany\\
{$^{4}$}Max Planck Institute for the Science of Light, Erlangen, Germany\\
$^*$emobi025@uottawa.ca, rasoul.alaee@mpl.mpg.de}

\date{\today}

\begin{abstract}
We theoretically investigate the optical force exerted on an isotropic particle illuminated by a superposition of plane waves. We derive explicit analytical expressions for the exerted force up to quadrupolar polarizabilities. Based on these analytical expressions, we demonstrate that an illumination consisting of two tilted plane waves can provide a full control on the optical force. In particular, optical pulling, pushing and lateral forces can be obtained by the proper tuning of illumination parameters. Our findings might unlock multiple applications based on a deterministic control of the spatial motion of small particles.


\end{abstract}


\maketitle


\section{Introduction}
Light as an electromagnetic radiation carries energy and momentum.
The direction of energy flux of an electromagnetic wave at any point
in space-time is given by the Poynting vector $\mathbf{S}(\mathbf{r},t)$ and the
linear momentum density (i.e. linear momentum per unit volume) is
given by $\mathbf{S}(\mathbf{r},t)/c^{2}$, where $c$ is the speed
of light~\cite{zangwill2013modern}. The exchange of linear momentum of light with an electromagnetically interacting particle can lead to an exerted optical force, known as
\textit{the radiation pressure.} In general, the acceleration caused by the radiation pressure on heavy macroscopic objects is considerably small. However, this acceleraion can be considerably large for small particles (compared to wavelength) when illuminated by a light beam with a moderate intensity. Therefore, light beams can be used to move, trap, or guide a particle. This became experimentally feasible after the invention of lasers and the challenge was overcome by Arthur Ashkin that used a single weakly focused laser beam/two counter-propagating beams to move/trap microparticles~\cite{ashkin1987optical,ashkin1970acceleration}. From then on, the technique has been widely used to manipulate atoms, molecules \cite{raab1987trapping,ashkin1978trapping}, and biological cells~\cite{welte1998developmental,ashkin1987optical}; and
it has opened a brand new field of research termed as optical manipulation~\cite{grier2003revolution,marago2013optical}.

Besides, one can have further control on the direction of the exerted optical
force and achieve counter-intuitive forces like the optical pulling
or lateral forces~\cite{chen2011optical,saenz2011optical,novitsky2011single,dogariu2013optically,brzobohaty2013,rodriguez2015lateral,AlaeePT2018}. The former being also called an optical tractor beam. These forces have been obtained by engineering the excitation and particle's symmetry and material. In particular,
optical pulling force on an isotropic particle can be obtained through
the interference of multiple plane waves, solenoidal beams~\cite{grier2014optical},
or Bessel beams~\cite{chen2015fano}. Furthermore, employing chiral
particles, gain media, or plasmonic interfaces can also allow achieving exerted lateral and pulling forces on the particle~\cite{webb2011negative,ding2014realization,novitsky2014pulling,wang2014lateral,canaguier2015chiral,Fernandes2015,Fernandes2016,Liu:17,petrov2016surface,AlaeePT2018}.

Multipole expansion is a key tool to study several optical phenomena namely, light perfect absorption~\cite{Landy:08,Alaee:16,alaee2017theory}, directional light emission~\cite{Steven:13,Hancu:13,Fu:13,Coenen:14,AlaeeKerker:15}, manipulating and controlling spontaneous emission~\cite{Rogobete:07,Zambrana:2015,Doeleman2016},  electromagnetically-induced-transparency~\cite{Chiam:09}, Fano resonances~\cite{Lukyanchuk:10,Miroshnichenko:10}, electromagnetic cloaking~\cite{Alu2008,Alu2009}, and also optical force~\cite{chaumet2000time,hayat2015lateral,guclu2015photoinduced,mobini2017optical,Albooyeh:2017,Kamandi:2017} and torque~\cite{PhysRevA.68.033802,nieto2015optical,chang1985optical,PhysRevB.94.125123} among many others. For small particles compared to the wavelength, induced electric and magnetic dipole and quadrupole moments are usually sufficient to fully understand the underlying physics. In this paper, we derive analytical expressions for the exerted optical force based on multipole expansion~\cite{bohren2008absorption,xu1995electromagnetic,barton1989theoretical,almaas1995radiation,rahimzadegan2017fundamental} up to quadrupolar polarizabilities. Detailed derivations are given in the \textit{supplementary material}. We theoretically and numerically study optical pulling, pushing, and lateral forces exerted on an isotropic particle for single and two tilted plane waves. The conditions for optical pushing, pulling, and lateral forces
are discussed. In particular, we explore the effects of the illumination
parameters i.e the wavelength, the angle between the two plane waves,
and the position of the particle on the exerted optical force. The
analytical expressions (\textit{Theory}) are verified with the numerical solution of
Maxwell's equations using COMSOL (\textit{Simulation}). The
electric and magnetic fields obtained through the simulations are employed
to calculate the Maxwell stress tensor and finally the optical force
can be calculated accordingly, see Eq.~(\ref{eq:a1}). In the following, we focus on the underlying theory and explain our results and its physical implications.

\section{Theory}

The time averaged mechanical force exerted on an arbitrary particle by an optical wave can be calculated as~\cite{jackson1999classical,novotny2012principles,zangwill2013modern}:

\begin{equation}
\left\langle \mathbf{F}\right\rangle =\left\langle \underset{s}{\varoint}{\underline{\mathbf{T}}}(\mathbf{r},t)\cdot\mathbf{n}\mathrm{d}S\right\rangle ,\label{eq:a1}
\end{equation}

\noindent with $S$ being any closed surface surrounding the particle,
$\mathbf{n}$  the outward unit normal vector to the surface, and
${\underline{\mathbf{T}}}$ the Maxwell's stress tensor.
The underline denotes the time-domain expressions. The Maxwell stress
tensor (MST) is a tensor of second rank whose elements are defined
as~\cite{jackson1999classical,novotny2012principles}:

\begin{equation}
\underline{T}_{ij}=\varepsilon_{0}\left[\underline{E}_{i}\underline{E}_{j}+c^{2}\underline{B}_{i}\underline{B}_{j}-\frac{1}{2}\delta_{ij}\left(|\underline{\mathbf{E}}|^{2}+|\underline{\mathbf{B}}|^{2}\right)\right]\label{eq:a2}
\end{equation}

\noindent where $\underline{E}$, $\underline{B}$ are the total (incoming plus scattered)
electric and magnetic fields in the $i,j=x,y,z$ axis, respectively.
$\delta_{ij}$ represents the Kronecker delta function.

\begin{figure}
\begin{centering}
\includegraphics[width=8.6cm]{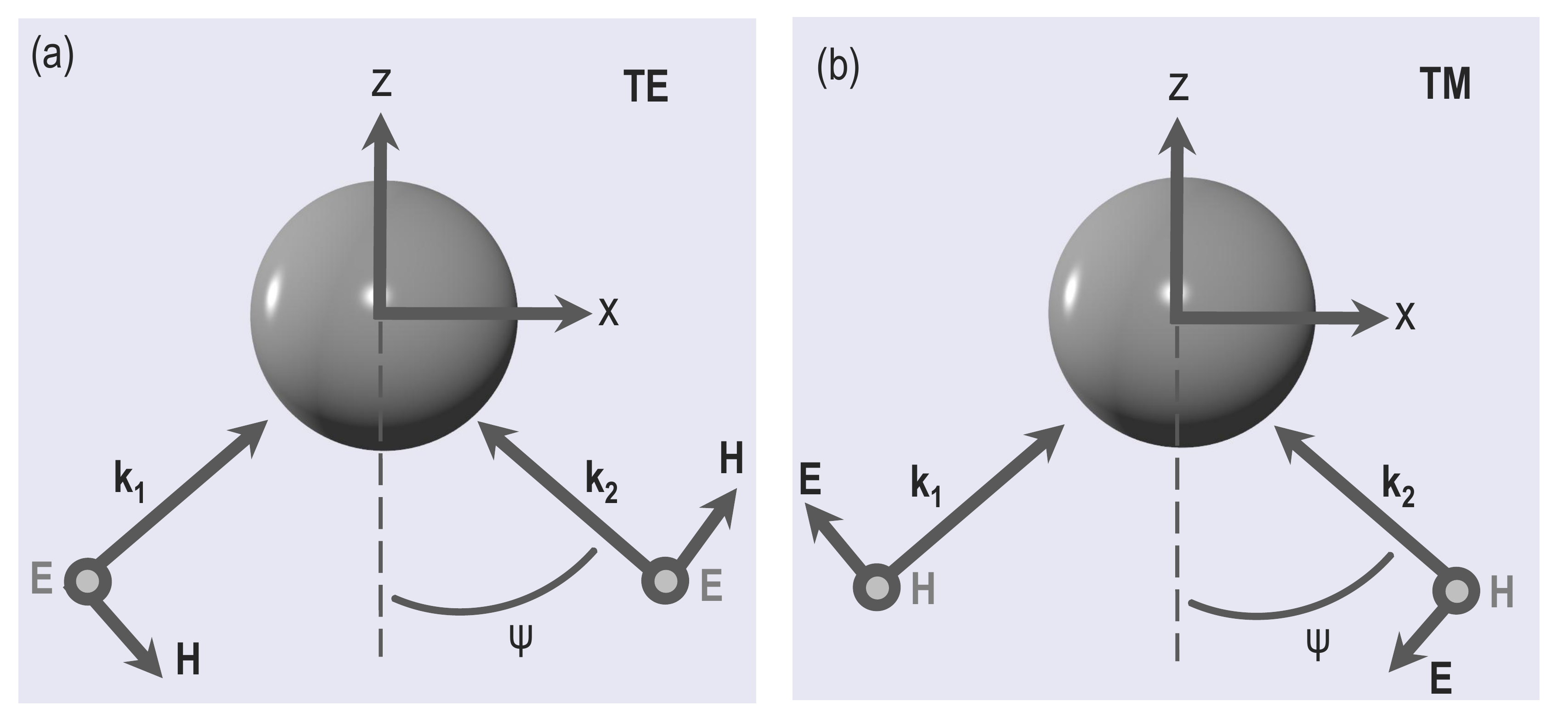}
\par\end{centering}

\caption{(a)-(b) The schematic of an isotropic particle illuminated by two tilted linearly polarized transverse electric (TE) and transverse magnetic (TM) plane waves, respectively. \label{fig:X1}}
\end{figure}

Although the method based on the MST provides the exact solution,
it does not suit the interpretative understanding of the light-matter
interaction. To this end, we use the multipole expansion method.
We decompose the induced current density in the particle in terms
of the electric and the magnetic multipole moments. For a small polarizable
particle (compared to wavelength), depending on the geometry, material, and the illumination,
the decomposition of induced current leads to the definition of the electric and magnetic
dipole and quadrupole polarizabilities (neglecting the higher order polarizabilities). The induced electric and
magnetic dipoles in Cartesian coordinates for an isotropic particle are expressed in terms of electric ($\alpha^{e}$) and magnetic ($\alpha^{m}$) polarizabilities
as $\mathbf{p}=\varepsilon_{0}\alpha^{e}\mathbf{E}$ and $\mathbf{m}=\alpha^{m}\mathbf{H}$,
respectively. The electric and magnetic quadrupole moments for an isotropic particle is defined as~\cite{Alu:2009,bernal2014underpinning}:

\begin{eqnarray*}
\mathbf{Q}^{e} & = & \varepsilon_{0}\alpha_{Q^{e}}\frac{\nabla\mathbf{E}+\mathbf{E}\nabla}{2},\\
\mathbf{Q}^{m} & = & \alpha_{Q^{m}}\frac{\nabla\mathbf{H}+\mathbf{H}\nabla}{2},
\end{eqnarray*}

\noindent where $\alpha_{Q^{e}}$ and $\alpha_{Q^{m}}$ are the Cartesian
quadrupolar polarizabilities. Using the relation $\left(\nabla\mathbf{A}+\mathbf{A}\nabla\right)_{ij}	=	\partial_{i}A_{j}+\partial_{j}A_{i}$,
any component of $Q^{e}$ and $Q^{m}$ is calculated as $Q_{ij}^{e}=\varepsilon_{0}\alpha_{Q^{e}}\left(\partial_{i}E_{j}+\partial_{j}E_{i}\right)/2$
and $Q_{ij}^{m}=\alpha_{Q^{m}}\left(\partial_{i}H_{j}+\partial_{j}H_{i}\right)/2$,
respectively.

The optical force exerted on a particle by an arbitrary illumination
can be written as the following (truncated at quadrupole order) \cite{barton1989theoretical,almaas1995radiation,rahimzadegan2017fundamental}:

\begin{eqnarray}
\mathbf{F} & = & \left[\mathbf{F}_{p}+\mathbf{F}_{Q^{e}}+...\right]+\left[\mathbf{F}_{m}+\mathbf{F}_{Q^{m}}+...\right]\nonumber \\
 &  & +\left[\mathbf{F}_{pm}+\mathbf{F}_{Q^{e}Q^{m}}+...\right] \label{eq:a4} \\
 & + & \left[\mathbf{F}_{pQ^{e}}+...\right]+\left[\mathbf{F}_{mQ^{m}}+...\right],\nonumber
\end{eqnarray}

\noindent where $\mathbf{F}_{p}(\mathbf{F}_{m})$ is the individual
contribution of an electric (magnetic) dipole moment to the total
optical force. $\mathbf{F}_{Q^{e}}(\mathbf{F}_{Q^{m}})$ is the contribution
of an individual electric (magnetic) quadrupole moment; and the other
expressions are the contribution of the interference between two multipole
moments.

\begin{widetext}
Neglecting higher order terms, Eq.~(\ref{eq:a4}) can be rewritten
in terms of Cartesian dipole and quadrupole moments as follows~\cite{barton1989theoretical,almaas1995radiation,chen2011optical,rahimzadegan2017fundamental}:

\begin{eqnarray}
F_{i} & = & \frac{1}{2}\textrm{Re}\left[\underset{j}{\sum}p_{j}\nabla_{i}E_{j}^{*}\right]+\frac{1}{2}\textrm{Re}\left[\underset{j}{\sum}m_{j}\nabla_{i}B_{j}^{*}\right]-\frac{k^{4}}{12\pi\varepsilon_{0}c}\textrm{Re}\left[\underset{j,k}{\sum}\epsilon_{ijk}p_{j}m_{k}^{*}\right]\nonumber \\
 & - & \frac{k^{5}}{120\pi\varepsilon_{0}}\textrm{Im}\left[\underset{j}{\sum}\left(Q^{e}\right)_{ij}p_{j}^{*}\right]+\frac{1}{\text{12}}\textrm{Re}\left[\left(Q^{e}\right)_{jk}\nabla_{i}\nabla_{k}E_{j}^{*}\right]+\frac{k^{5}}{120\pi\varepsilon_{0}c^{2}}\textrm{Im}\left[\left(Q^{m}\right)_{ij}m_{j}^{*}\right]\nonumber \\
 & + & \frac{1}{12}\mathrm{Re}\left[\underset{k}{\sum}\left(Q^{m}\right)_{jk}\nabla_{i}\nabla_{k}B_{j}^{*}\right]-\frac{k^{6}}{9\times240\pi\varepsilon_{0}c}\mathrm{Re}\left[\underset{l,j,k}{\sum}\varepsilon_{ijk}(Q^{e})_{lj}\left(Q^{m}\right)_{lk}^{*}\right],\label{eq:a6}
\end{eqnarray}

\noindent in which $E_i$ and $B_i$ are components of the incident electric and magnetic
fields, respectively, and $\epsilon_{ijk}$ is the \textit{Levi-Civita}
symbol.

As shown in Fig.~\ref{fig:X1}, consider an illumination composed of two tilted plane waves with wave vectors $\mathrm{\mathbf{k}}_{1}=k[\mathrm{sin\psi},0,\mathrm{cos\psi}]$
and $\mathrm{\mathbf{k}}_{2}=k[-\mathrm{sin\psi},0,\mathrm{cos\psi}]$, illuminating an isotropic particle, where $\psi$ is the tilting angle. We define the TE and TM illuminations as the following:
\begin{eqnarray}
\mathbf{E^{\mathrm{TE}}} & = & \frac{E_{0}}{2}\left(e^{i\mathbf{k}_{1}\cdot\mathbf{r}}+e^{i\mathbf{k}_{2}\cdot\mathbf{r}}\right)\mathbf{e}_{y},\label{eq:TE}\\
\mathrm{\mathbf{E}}^{\mathrm{TM}} & = & \frac{E_{0}}{2}\left(\left[\begin{array}{c}
\mathrm{cos}\psi\\
0\\
-\mathrm{sin}\psi
\end{array}\right]e^{i\mathbf{k}_{1}\cdot\mathbf{r}}+\left[\begin{array}{c}
\mathrm{cos}\psi\\
0\\
\mathrm{sin}\psi
\end{array}\right]e^{i\mathbf{k}_{2}\cdot\mathbf{r}}\right).\label{eq:TM}
\end{eqnarray}

The time averaged optical forces exerted on an isotropic particle located at the position $\mathbf{r}_{0}=\left(x_{0},y_{0},z_{0}\right)$ by the TE and TM
illuminations, i.e. Eqs.~(\ref{eq:TE})-(\ref{eq:TM}) read as (see supplementary material):


\begin{eqnarray}
\mathbf{\overline{F}}^{\mathrm{TE}} & \approx & \mathbf{\overline{F}}_{p}^{\mathrm{TE}}+\mathbf{\overline{F}}_{m}^{\mathrm{TE}}+\overline{\mathbf{F}}_{pm}^{\mathrm{TE}}+\mathbf{\overline{F}}_{Q^{e}}^{\mathrm{TE}}+\mathbf{\overline{F}}_{Q^{m}}^{\mathrm{TE}}+\mathbf{\overline{F}}_{pQ^{e}}^{\mathrm{TE}}+\mathbf{\overline{F}}_{mQ^{m}}^{\mathrm{TE}}+\mathbf{\overline{F}}_{Q^{e}Q^{m}}^{\mathrm{TE}},\label{eq:a8}\\
\overline{\mathbf{F}}_{p}^{\mathrm{TE}} & = & -\frac{3}{2}\mathrm{sin}\psi\mathrm{sin}2\delta\mathrm{Re}\left(\bar{\alpha_{e}}\right)\mathbf{e}_{x}+3\mathrm{cos}\psi\mathrm{cos^{2}}\delta\mathrm{Im}\left(\bar{\alpha_{e}}\right)\mathbf{e}_{z},\nonumber \\
\overline{\mathbf{F}}_{m}^{\mathrm{TE}} & = & -\frac{3}{2}\mathrm{sin}\psi\mathrm{sin}2\delta\mathrm{cos2\psi}\mathrm{Re}\left(\bar{\alpha}_{m}\right)\mathbf{e}_{x}+3\mathrm{cos}\psi\left(\mathrm{cos^{2}\psi cos^{2}}\delta+\mathrm{sin^{2}}\psi\mathrm{sin^{2}}\delta\right)\mathrm{Im}\left(\bar{\alpha_{m}}\right)\mathbf{e}_{z},\nonumber \\
\overline{\mathbf{F}}_{pm}^{\mathrm{TE}} & = & -\frac{3}{2}\mathrm{sin\psi}\mathrm{sin}2\delta\mathrm{Im}\left(\bar{\alpha_{e}}\bar{\alpha}_{m}^{*}\right)\mathbf{e}_{x}-3\mathrm{cos}\psi\mathrm{cos^{2}}\delta\mathrm{Re}\left(\bar{\alpha_{e}}\bar{\alpha}_{m}^{*}\right)\mathbf{e}_{z},\nonumber \\
\overline{\mathbf{F}}_{Q^{e}}^{\mathrm{TE}} & = & -\frac{5}{2}\mathrm{sin}\psi\mathrm{sin}2\delta\mathrm{cos}2\psi\mathrm{Re}\left(\bar{\alpha}_{Q^{e}}\right)\mathbf{e}_{x}+5\mathrm{cos}\psi\left(\mathrm{cos^{2}\psi cos^{2}}\delta+\mathrm{sin^{2}}\psi\mathrm{sin^{2}}\delta\right)\mathrm{Im}\left(\bar{\alpha}_{Q^{e}}\right)\mathbf{e}_{z},\nonumber \\
\overline{\mathbf{F}}_{Q^{m}}^{\mathrm{TE}} & = & -\frac{5}{2}\mathrm{sin}\psi\mathrm{sin}2\delta\mathrm{cos4\psi}\mathrm{Re}\left(\bar{\alpha}_{Q^{m}}\right)\mathbf{e}_{x}+5\mathrm{cos}\psi\left(\mathrm{cos^{2}2\psi cos^{2}}\delta+\mathrm{sin^{2}2\psi sin^{2}}\delta\right)\mathrm{Im}\left(\bar{\alpha}_{Q^{m}}\right)\mathbf{e}_{z},\nonumber \\
\overline{\mathbf{F}}_{pQ^{e}}^{\mathrm{TE}} & = & +\frac{3}{2}\mathrm{sin}\psi\mathrm{sin}2\delta\mathrm{Im\left(\bar{\alpha}_{Q^{e}}\bar{\alpha}_{e}^{*}\right)}\mathbf{e}_{x}-3\mathrm{cos}\psi\mathrm{cos^{2}}\delta\mathrm{Re}\left(\bar{\alpha}_{Q^{e}}\bar{\alpha}_{e}^{*}\right)\mathbf{e}_{z},\nonumber \\
\overline{\mathbf{F}}_{mQ^{m}}^{\mathrm{TE}} & = & +\frac{3}{2}\mathrm{sin}\psi\mathrm{sin}2\delta\left(\mathrm{cos2\psi+2cos^{2}\psi}\right)\mathrm{Im}\left(\bar{\alpha}_{Q^{m}}\bar{\alpha}_{m}^{*}\right)\mathbf{e}_{x}-3\mathrm{cos}\psi\left(\mathrm{cos}2\psi\mathrm{cos^{2}}\delta+2\mathrm{sin^{2}}\psi\mathrm{sin^{2}}\delta\right)\mathrm{Re}\left(\bar{\alpha}_{Q^{m}}\bar{\alpha}_{m}^{*}\right)\mathbf{e}_{z},\nonumber \\
\overline{\mathbf{F}}_{Q^{e}Q^{m}}^{\mathrm{TE}} & = & -\frac{5}{6}\mathrm{sin\psi\mathrm{sin}2\delta\mathrm{\left(cos2\psi+2cos^{2}\psi\right)}Im}\left(\bar{\alpha}_{Q^{e}}\bar{\alpha}_{Q^{m}}^{*}\right)\mathbf{e}_{x}-\frac{5}{3}\mathrm{cos}\psi\left(\mathrm{cos}2\psi\mathrm{cos^{2}}\delta+2\mathrm{sin^{2}\psi sin^{2}}\delta\right)\mathrm{Re}\left(\bar{\alpha}_{Q^{e}}\bar{\alpha}_{Q^{m}}^{*}\right)\mathbf{e}_{z},\nonumber
\end{eqnarray}


\begin{eqnarray}
\overline{\mathbf{F}}^{\mathrm{TM}} & \approx & \mathbf{\overline{F}}_{p}^{\mathrm{TM}}+\mathbf{\overline{F}}_{m}^{\mathrm{TM}}+\mathbf{\overline{F}}_{pm}^{\mathrm{TM}}+\mathbf{\overline{F}}_{Q^{e}}^{\mathrm{TM}}+\mathbf{\overline{F}}_{Q^{m}}^{\mathrm{TM}}+\mathbf{\overline{F}}_{pQ^{e}}^{\mathrm{TM}}+\mathbf{\overline{F}}_{mQ^{m}}^{\mathrm{TM}}+\mathbf{\overline{F}}_{Q^{e}Q^{m}}^{\mathrm{TM}},\label{eq:a9}\\
\mathbf{\overline{F}}_{p}^{\mathrm{TM}} & = & -\frac{3}{2}\mathrm{sin\psi}\mathrm{sin}2\delta\mathrm{cos2}\psi\mathrm{Re}\left(\bar{\alpha_{e}}\right)\mathbf{e}_{x}+3\mathrm{cos}\psi\mathrm{\left(\mathrm{cos^{2}\psi cos^{2}}\delta+\mathrm{sin^{2}}\psi\mathrm{sin^{2}}\delta\right)\mathrm{Im}\left(\bar{\alpha_{e}}\right)}\mathbf{e}_{z},\nonumber \\
\mathbf{\overline{F}}_{m}^{\mathrm{TM}} & = & -\frac{3}{2}\mathrm{sin\psi}\mathrm{sin}2\delta\mathrm{Re}\left(\bar{\alpha_{m}}\right)\mathbf{e}_{x}+3\mathrm{cos}\psi\mathrm{cos^{2}}\delta\mathrm{Im}\left(\bar{\alpha_{m}}\right)\mathbf{e}_{z},\nonumber \\
\mathbf{\overline{F}}_{pm}^{\mathrm{TM}} & = & +\frac{3}{2}\mathrm{sin\psi}\mathrm{sin}2\delta\mathrm{Im}\left(\bar{\alpha_{e}}\bar{\alpha}_{m}^{*}\right)\mathbf{e}_{x}-3\mathrm{cos}\psi\mathrm{cos^{2}}\delta\mathrm{Re}\left(\bar{\alpha_{e}}\bar{\alpha}_{m}^{*}\right)\mathbf{e}_{z},\nonumber \\
\mathbf{\overline{F}}_{Q^{e}}^{\mathrm{TM}} & = & -\frac{5}{2}\mathrm{sin}\psi\mathrm{sin}2\delta\mathrm{cos}4\psi\mathrm{Re}\left(\bar{\alpha}_{Q^{e}}\right)\mathbf{e}_{x}+5\mathrm{cos}\psi\left(\mathrm{cos^{2}2\psi cos^{2}}\delta+\mathrm{sin^{2}2\psi sin^{2}}\delta\right)\mathrm{Im}\left(\bar{\alpha}_{Q^{e}}\right)\mathbf{e}_{z},\nonumber \\
\mathbf{\overline{F}}_{Q^{m}}^{\mathrm{TM}} & = & -\frac{5}{2}\mathrm{sin}\psi\mathrm{sin}2\delta\mathrm{cos2\psi}\mathrm{Re}\left(\bar{\alpha}_{Q^{m}}\right)\mathbf{e}_{x}+5\mathrm{cos}\psi\left(\mathrm{cos^{2}\psi cos^{2}}\delta+\mathrm{sin^{2}}\psi\mathrm{sin^{2}}\delta\right)\mathrm{Im}\left(\bar{\alpha}_{Q^{m}}\right)\mathbf{e}_{z},\nonumber \\
\mathbf{\overline{F}}_{pQ^{e}}^{\mathrm{TM}} & = & +\frac{3}{2}\mathrm{sin}\psi\mathrm{sin}2\delta\left(\mathrm{cos2\psi+2cos^{2}\psi}\right)\mathrm{Im}\left(\bar{\alpha}_{Q^{e}}\bar{\alpha}_{e}^{*}\right)\mathbf{e}_{x}-3\mathrm{cos}\psi\left(\mathrm{cos^{2}}\delta\mathrm{cos}2\psi+2\mathrm{sin^{2}}\psi\mathrm{sin^{2}}\delta\right)\mathrm{Re}\left(\bar{\alpha}_{Q^{e}}\bar{\alpha}_{e}^{*}\right)\mathbf{e}_{z},\nonumber \\
\mathbf{\overline{F}}_{mQ^{m}}^{\mathrm{TM}} & = & +\frac{3}{2}\mathrm{sin}\psi\mathrm{sin}2\delta\mathrm{Im}\left(\bar{\alpha}_{Q^{m}}\bar{\alpha}_{m}^{*}\right)\mathbf{e}_{x}-3\mathrm{cos}\psi\mathrm{cos^{2}}\delta\mathrm{Re}\left(\bar{\alpha}_{Q^{m}}\bar{\alpha}_{m}^{*}\right)\mathbf{e}_{z},\nonumber \\
\overline{\mathbf{F}}_{Q^{e}Q^{m}}^{\mathrm{TM}} & = & +\frac{5}{6}\mathrm{sin}\psi\mathrm{sin}2\delta\left(\mathrm{\mathrm{cos2\psi+2cos^{2}\psi}}\right)\mathrm{Im}(\bar{\alpha}_{Q^{e}}\bar{\alpha}_{Q^{m}}^{*})\mathbf{e}_{x}-\frac{5}{3}\mathrm{cos}\psi\left(\mathrm{cos^{2}}\delta\mathrm{cos}2\psi+2\mathrm{sin^{2}\psi sin^{2}}\delta\right)\mathrm{Re}\left(\bar{\alpha}_{Q^{e}}\bar{\alpha}_{Q^{m}}^{*}\right)\mathbf{e}_{z}.\nonumber
\end{eqnarray}

\end{widetext}

\noindent where we define $\bar{\alpha}_{e,m}=\alpha_{e,m}/\alpha_{d}$,
$\bar{\alpha}_{Q^{e,m}}=\alpha_{Q^{e,m}}/\alpha_{q}$, and $\delta=k\mathrm{sin\psi}x_{0}$. $\alpha_{q}=120\pi/k^{5}$ and $\alpha_{d}=6\pi/k^{3}$ are the
polarizability normalizations for dipoles and quadrupoles, respectively. Please note that a lateral change in $x_0$, being the spatial position of the particle, has the equivalent effect on the force as a phase shift $\Delta\phi=-2kx_0\sin\psi$ in the two plane waves illuminating the particle. This makes sense as the spatial interference pattern that the two plane waves form only depends on the phase difference.
The optical force exerted on the particle only depends on its position along the $x$-axis while it is independent on the position along the $z$- and $y$-axis (i.e. it depends on $\delta=k\mathrm{sin\psi}x_{0}$).
Throughout the paper, optical forces are normalized to $\mathrm{F}^{\mathrm{norm}}=\left(I_{0}/c\right)\left[\lambda^{2}/\left(2\pi\right)\right]=I_{0}k\alpha_{d}/\left(3c\right)$.
The normalization factor of $\mathrm{F}^{\mathrm{norm}}$ is of physical significance
and $\mathrm{3F}^{\mathrm{norm}}$ corresponds to the upper bound for the exerted
optical force on an isotropic electric/magnetic dipolar particle illuminated by a plane wave~\cite{rahimzadegan2017fundamental}.

\section{Theoretical and numerical results}
\begin{figure}
\begin{centering}
\includegraphics[width=8.6cm]{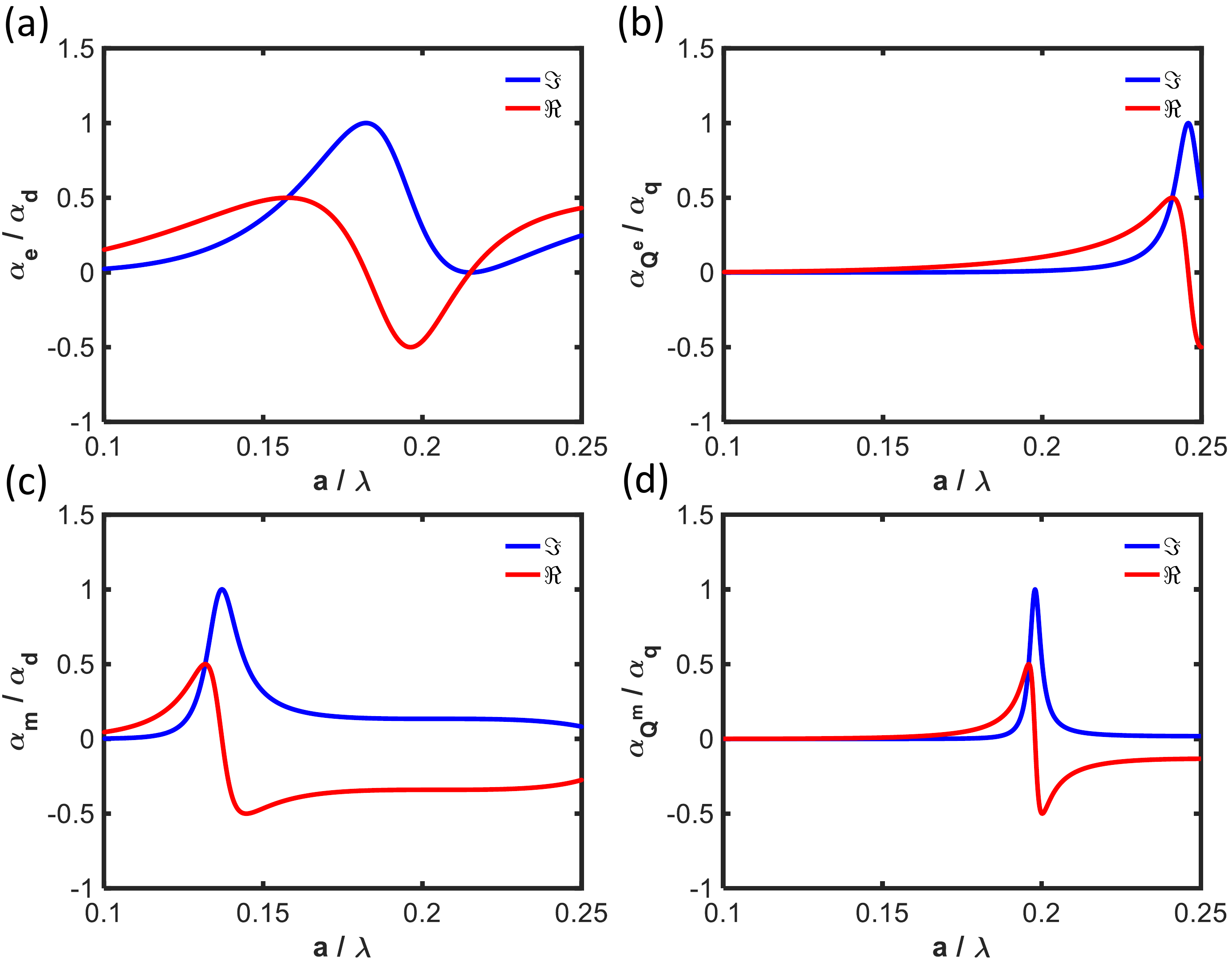}
\par\end{centering}

\caption{(a)-(b) Real and imaginary parts of the normalized Cartesian dipole and quadrupole
polarizabilities of a sphere with permittivity
$\varepsilon=3.5^{2}$
and radius $a$. \label{fig:Polarizability}}
\end{figure}

In the following, we consider a dielectric sphere made from a material with a permittivity of
$\varepsilon=3.5^{2}$, and radius $a$. Figure~\ref{fig:Polarizability}
shows the calculated polarizabilities. They are calculated by using electric and magnetic Mie coefficients (i.e. $a_1,a_2$ and $b_1,b_2$)~\cite{mie1908beitrage}:
\begin{eqnarray}
\alpha_{e}=i\frac{6\pi}{k^{3}}a_{1}=i\alpha_{d}a_{1}, & \,\,\,\,\, & \alpha_{m}=i\frac{6\pi}{k^{3}}b_{1}=i\alpha_{d}b_{1},\\
\alpha_{Q^{e}}=i\frac{120\pi\epsilon_{0}}{k^{5}}a_{2}=i\alpha_{q}a_{2}, & \,\,\,\,\, & \alpha_{Q^{m}}=i\frac{120\pi}{k^{5}}b_{2}=i\alpha_{q}b_{2}.\nonumber
\end{eqnarray}
Alternatively, they can be extracted from exact multipole moments based on induced current~\cite{FernandezCorbaton:15,Alaee:2018}. We restrict ourselves to the wavelength region in where the lowest order polarizabilities have their lowest order resonances. Having the polarizabilities, through Eqs.~(\ref{eq:a8}) and (\ref{eq:a9}), the exerted optical force due to the contribution of different multipole moments can be derived. Below, we consider several scenarios for the illumination.

\subsection{Single plane wave illumination}

Assuming a plane wave excitation, i.e. $\psi=0$, using Eqs.(\ref{eq:a8}) or (\ref{eq:a9}) the
normalized optical force exerted on an isotropic particle is calculated
as:

\begin{eqnarray}
\overline{\mathbf{F}} & \approx & 3\mathrm{Im}\left(\bar{\alpha}_{e}+\bar{\alpha}_{m}\right)+5\mathrm{Im}\left(\bar{\alpha}_{Q^{e}}+\bar{\alpha}_{Q^{m}}\right)\label{eq:PW_EP}\\
 & - & 3\mathrm{Re}\left(\bar{\alpha}_{e}\bar{\alpha}_{m}^{*}+\bar{\alpha}_{Q^{e}}\bar{\alpha}_{e}^{*}+\bar{\alpha}_{Q^{m}}\bar{\alpha}_{m}^{*}+\frac{5}{9}\bar{\alpha}_{Q^{e}}\bar{\alpha}_{Q^{m}}^{*}\right)\mathbf{e}_z.\nonumber
\end{eqnarray}

\begin{figure}
\begin{centering}
\includegraphics[width=8.6cm]{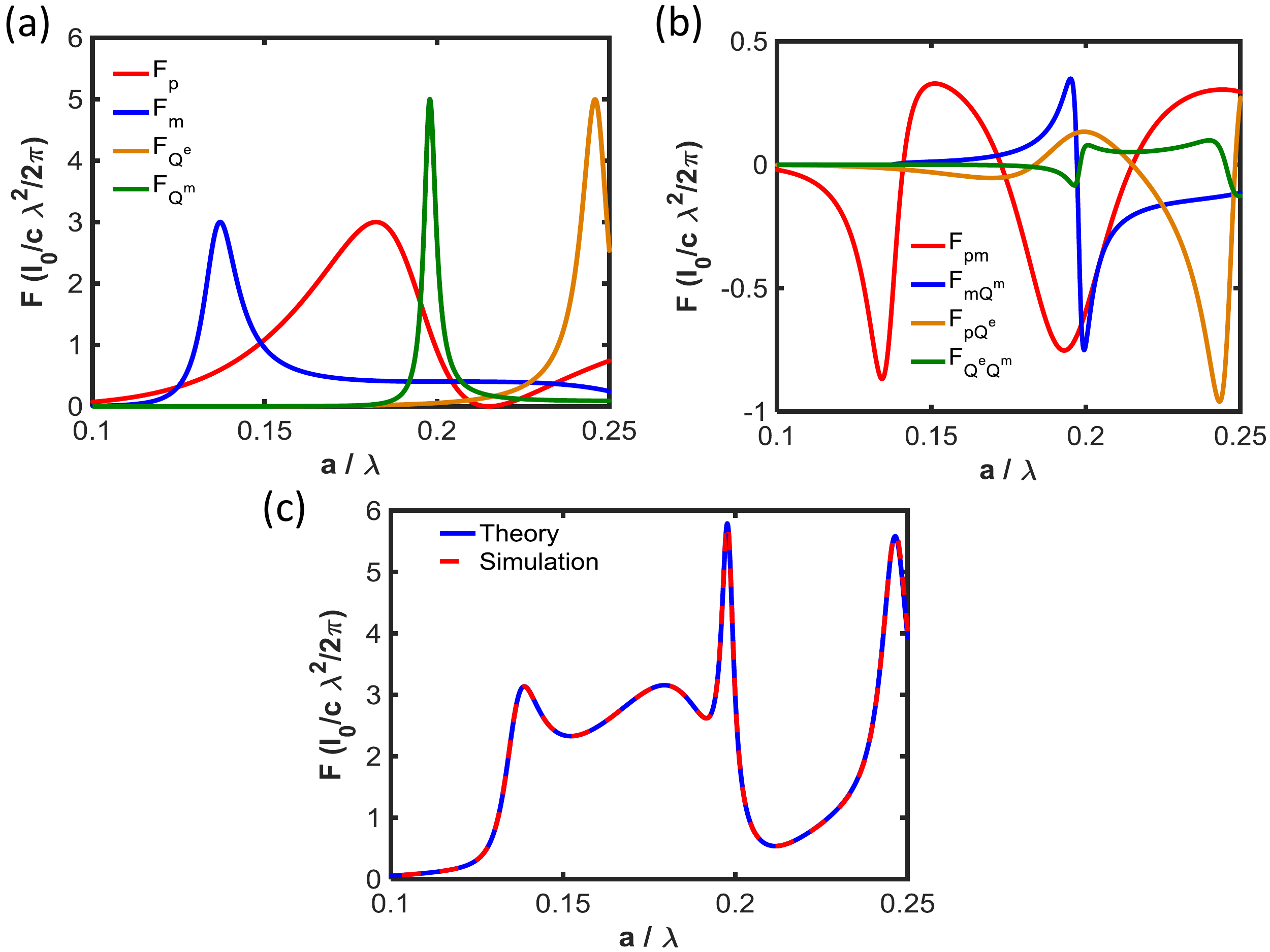}
\par\end{centering}

\caption{\textit{Single plane wave illumination}: (a)-(b) Contribution of different moments to the multipolar force exerted
by a single plane wave. (c) The exerted analytical and simulated optical force calculated by the analytical expression, i.e. Eq.~(\ref{eq:PW_EP}) and the numerical approach based on MST, i.e. Eq.~(\ref{eq:a1}), respectively.\label{fig:Force_1PW}}
\end{figure}

The theoretical results based on the derived equation are shown in Fig.~\ref{fig:Force_1PW}. As a verification, the results of the COMSOL simulation are also shown in Fig.~\ref{fig:Force_1PW}(c).

As can be seen in Fig.~\ref{fig:Force_1PW}(a) and (b), interference terms can have negative (pulling) contributions to the total optical force, while the individual contributions
of the moments are positive across the entire spectrum (the imaginary
part of the polarizabilities is always positive for passive particles).
With a single plane wave illumination, the total optical force is always positive (pushing), Fig.~\ref{fig:Force_1PW}(c). Further, it can be observed that the quadrupolar terms have the dominant contributions to the highest values of the optical force at lower wavelengths. The optical
force, as can be intuitively expected, is in the direction of the overall
linear momentum, i.e. here, in the $z$-direction.

\subsection{Two plane wave illumination: sphere at $\mathbf{r}=\mathbf{0}$}

\begin{figure}
\begin{centering}
\includegraphics[width=8.6cm]{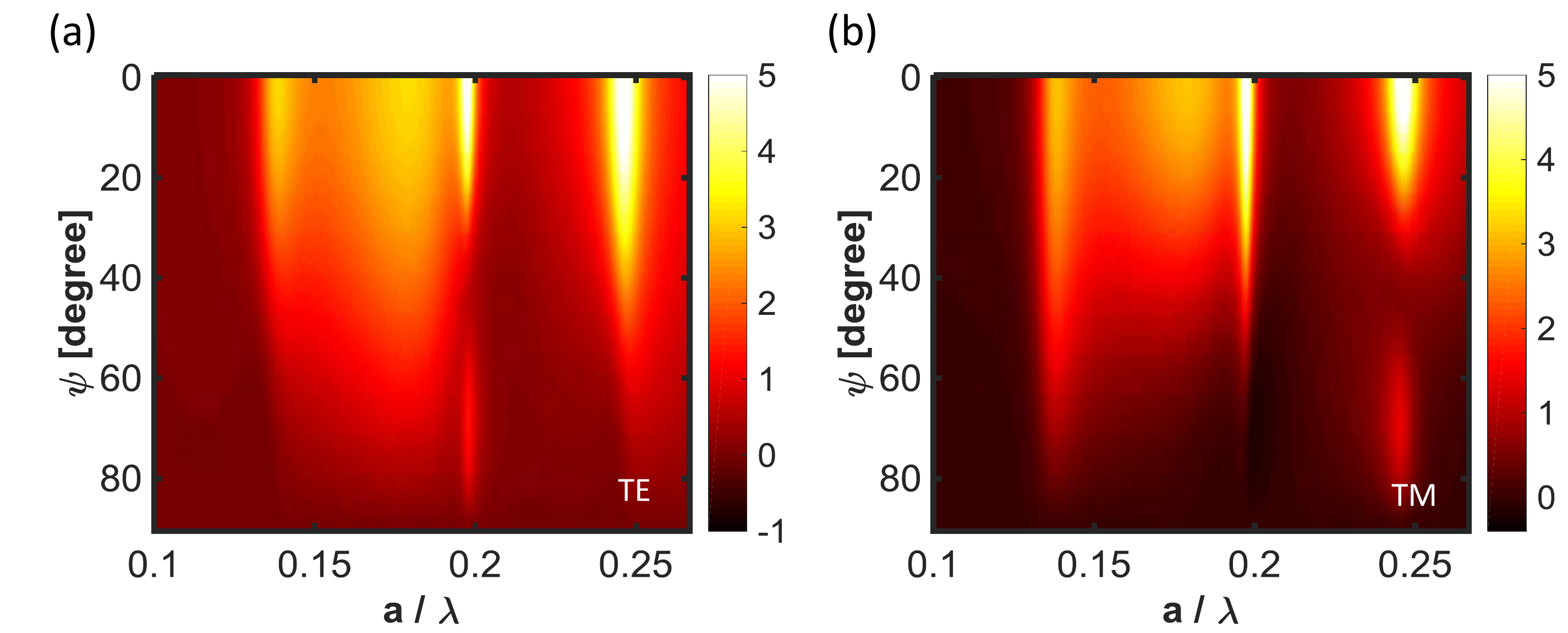}
\par\end{centering}

\caption{\textit{Two plane wave illumination and the sphere located at $\mathbf{r}=0$}: (a)-(b) Exerted optical forces by two tilted plane waves as a function of
the tilting angle and the particle's size parameter, i.e. $a/\lambda$ for TE and
TM illuminations, respectively.\label{fig:Force_2PW}}
\end{figure}

Assuming the sphere to be located in the center of the coordinate system, i.e $\mathbf{r}=(0,0,0)$,
and being illuminated with the wave expressed by Eqs.~\ref{eq:TE}-\ref{eq:TM},
the calculated optical force as a function of the tilting angle $\psi$ and the particle's size parameter, i.e. $a/\lambda$, is shown in Fig.~\ref{fig:Force_2PW}(a) and (b) for TE- and TM-polarization, respectively.
An optical pulling force is achieved for certain tilting angles for both
TE and TM illuminations. This can be expected already by inspecting
Eqs.~(\ref{eq:a8})-(\ref{eq:a9}), since the contribution of negative
terms can dominate for some large angles as the positive terms
are attenuating faster as the angle increases.\\
To make a better analysis of the negative force,
we choose a smaller sized sphere, where dipole moments are dominant,
and neglect the quadrupolar moments. Therefore, the optical force
simplifies to:
\begin{eqnarray}
\overline{\mathbf{F}} & \approx & \overline{\mathbf{F}}_{p}+\overline{\mathbf{F}}_{m}+\overline{\mathbf{F}}_{pm},\label{eq:Dipole_F}\\
\mathrm{\overline{\mathbf{F}}^{TE}} & \approx & 3\mathrm{cos}\psi\left[\mathrm{Im}\left(\bar{\alpha}_{e}\right)+\mathrm{Im}\left(\bar{\alpha}_{m}\right)\mathrm{cos^{2}}\psi-\mathrm{Re}\left(\bar{\alpha}_{e}\bar{\alpha}_{m}^{*}\right)\right]\mathbf{e}_{z},\nonumber \\
\mathrm{\overline{\mathbf{F}}^{TM}} & \approx & 3\mathrm{cos}\psi\left[\mathrm{Im}\left(\bar{\alpha}_{e}\right)\mathrm{cos^{2}}\psi+\mathrm{Im}\left(\bar{\alpha}_{m}\right)-\mathrm{Re}\left(\bar{\alpha}_{e}\bar{\alpha}_{m}^{*}\right)\right]\mathbf{e}_{z}.\nonumber
\end{eqnarray}

\begin{figure}
\begin{centering}
\includegraphics[width=8.6cm]{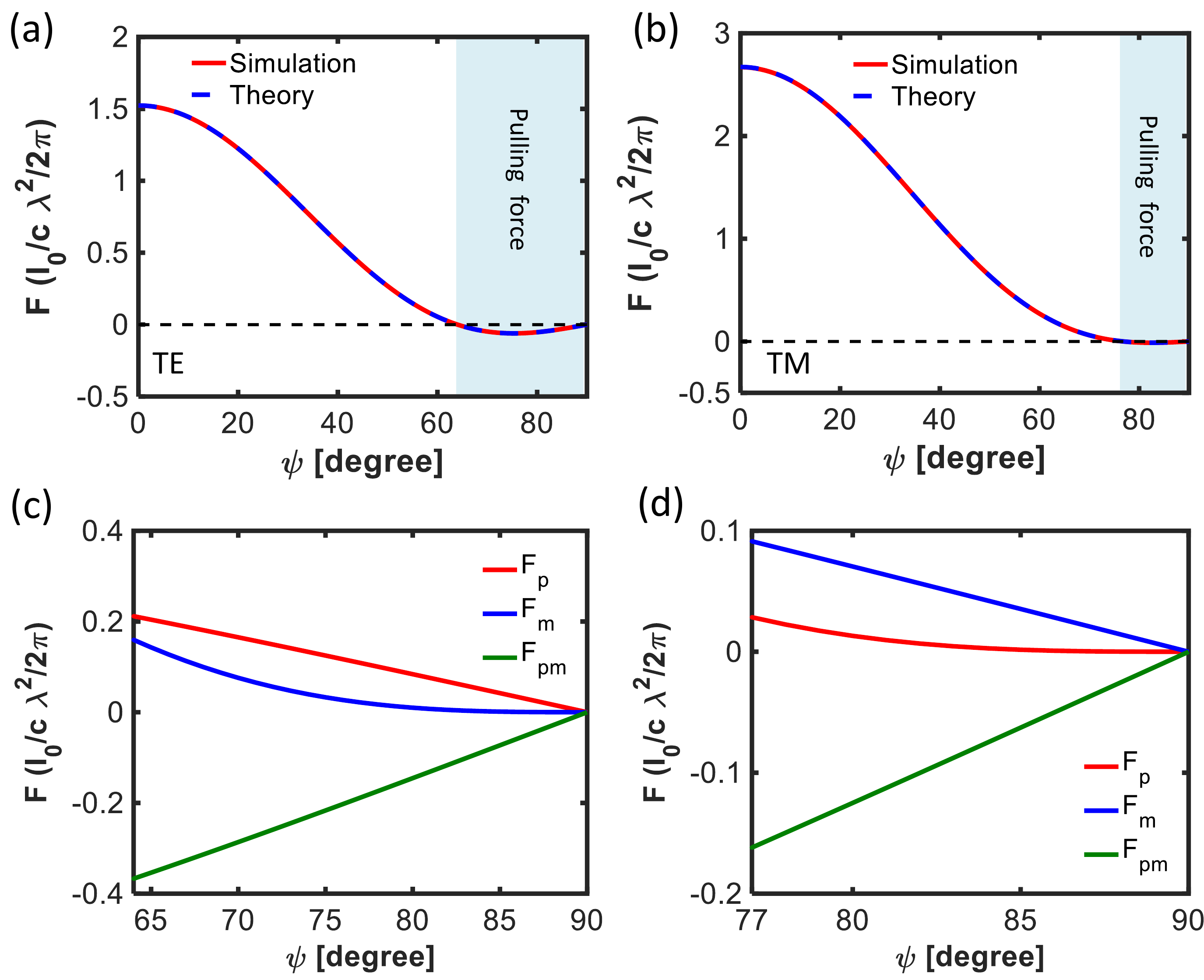}
\par\end{centering}

\caption{\textit{Two plane wave illumination and the sphere located at $\mathbf{r}=0$}: (a)-(b) Exerted optical forces on the dielectric sphere, with permittivity
$\varepsilon=3.5^{2}$, by two tilted plane waves
at $a/\lambda=\left(1/7.52\right)$ and  $\left(1/5.26\right)$ for TE and TM illuminations, respectively. (c)-(d) Contribution of different orders of dipoles and their interference
to the optical force.\label{fig:dipole_contribution}}
\end{figure}

The calculated optical force is shown in Fig.~\ref{fig:dipole_contribution}
for certain values of $a/\lambda=\left(1/7.52\right)$, $\left(1/5.26\right)$
for TE and TM illuminations, respectively.

\begin{figure}
\begin{centering}
\includegraphics[width=8.6cm]{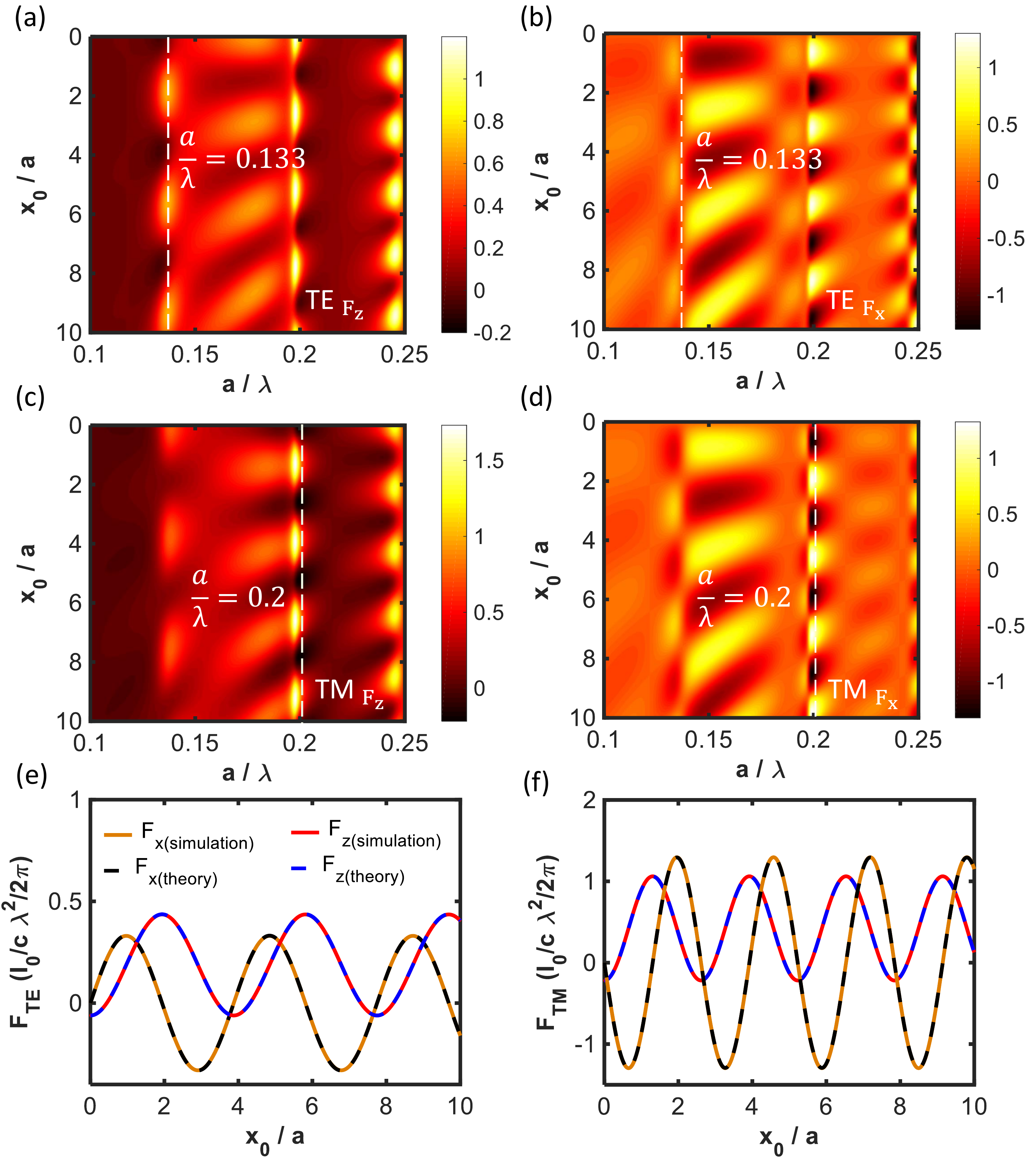}
\par\end{centering}

\caption{\textit{Two plane wave illumination and the sphere located at $\mathbf{r}=\mathbf{r}_0$}: (a-d) Exerted optical forces on the dielectric sphere as a function of its position on the $x$-axis and its size parameter, i.e. $a/\lambda$ for TE and TM illuminations. (e,f) Periodic optical pushing/pulling forces on the dielectric sphere at $a/\lambda=0.133$ and $0.2$ for both TE and TM illuminations, respectively.  \label{fig:periodic_force}}
\end{figure}

\begin{figure*}
\begin{centering}
\includegraphics[width=15cm]{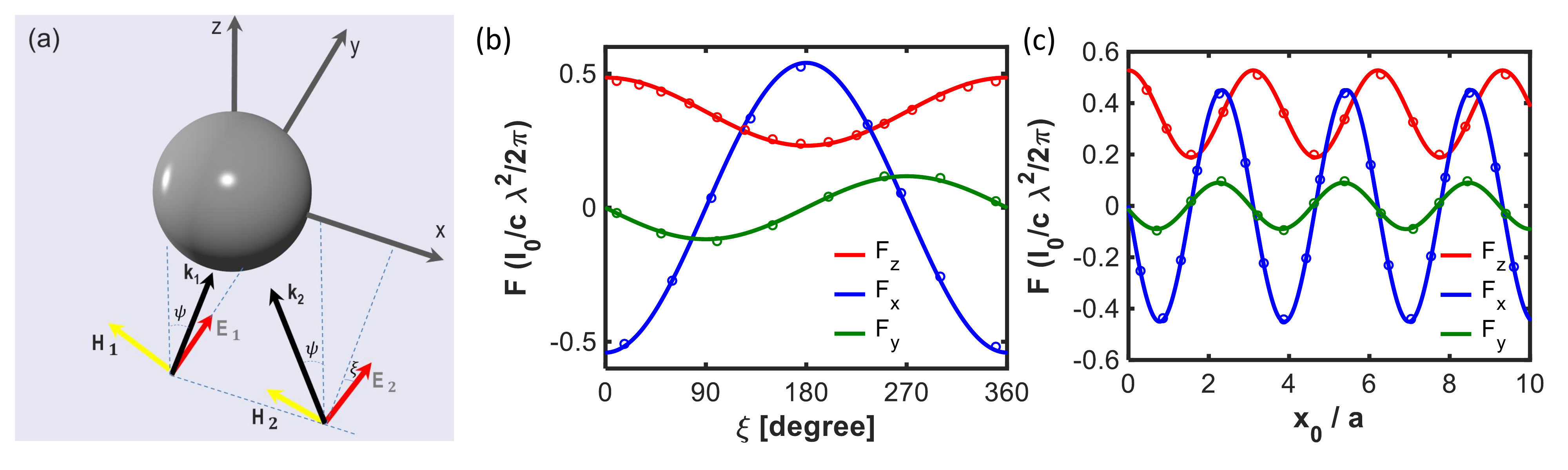}
\par\end{centering}

\caption{\textit{Two plane wave illumination using Eq.~(\ref{eq:PWI}) and sphere located at $\mathbf{r}=\mathbf{r}_0$}: (a) The schematic of an isotropic particle illuminated by two tilted linearly polarized plane waves. (b)-(c) Optical forces versus $\xi$ and $x_0/a$ exerted on the dielectric sphere with permittivity $\varepsilon=3.5^{2}$ and illuminated by two tilted linearly polarized plane waves at wavelength $\lambda=6\,a$ with angles ($\psi=75^{\circ}$,
$x_0=a/2$). (c) for $\psi=75^{\circ},$ $\xi=45^{\circ}$. The solid line is the simulation and the circles are the analytical results calculated by using Eq.~(\ref{eq:a11}). \label{fig:angle_variation}}
\end{figure*}

%

Based on this figure, for small values of the tilting angle $\psi$,
a pushing force is exerted on the particle. As the tilting angle increases, for the TE(TM) polarization the magnitude of the force reduces until it vanishes at $\psi=62^{{\circ}}\left(76^{{\circ}}\right)$ and beyond that the pulling force appears, showing a minimum at $\psi=75^{\circ}\left(83^{{\circ}}\right)$.
For both TE and TM illuminations, the terms $F_{p}$ and $F_{m}$ are
positive, however, the term $F_{pm}$ in both cases is negative, canceling
out the contributions of the positive terms at certain angles. Then, it becomes possible to reduce the positive contributions to the optical force and to achieve an overall negative force. In other words, according to Eq.~(\ref{eq:Dipole_F}) for TE (TM) polarization, the term $\overline{\mathbf{F}}_m$ ($\overline{\mathbf{F}}_p$) vanishes due to the term $\mathrm{cos^3}\psi$ for large $\psi$ ~[see Fig.~\ref{fig:dipole_contribution} (a)-(d)]. Meanwhile, the overall force is decreased due to the term $\mathrm{cos}\psi$, a factor which appears as a total pre-factor.

\subsection{Two plane wave illumination: sphere at $\mathbf{r}=\mathbf{r}_0$}

In order to investigate the effects of the particle position $\mathbf{r}_{0}$ on the force exerted on it, which shows
itself in the equations through the angle $\delta=k\mathrm{sin\psi}x_{0}$,
the exerted optical forces as a function of the position and the particle's size parameter, i.e. $a/\lambda$ is shown in Fig.~\ref{fig:periodic_force} (a)-(d). 
The tilting angles for the TE and TM polarizations are $76^{\circ}$ and $73^{{\circ}}$,
respectively. It can be seen that the particle experiences a periodic optical force. The existence of the lateral force is due to the gradient of the field intensity
along the $x$-axis. Moreover, according to these figures it can be realized that the quadrupolar terms (around $a/\lambda = 0.2$ and $0.25$, see also Figure~\ref{fig:Polarizability}) cause major variations of the optical force in amplitude and sign for both TE and TM illuminations. Figure~\ref{fig:periodic_force} (e)-(f) depict the theoretical and simulated exerted forces (i.e. both $x$ and $z$ components) calculated for $a/\lambda = 0.133$ and $0.2$ with the periodicity of $\Lambda=\lambda/\mathrm{sin}\psi
 $ with respect to the x-axis (see the definition of $\delta$). Theoretical results using Eq.~(\ref{eq:a9}) are in perfect agreement with the simulation results.


In order to explore other possible influences of two-plane wave illumination in the $x$-$z$-plane on the exerted optical force, the following generalized excitation is defined:

\begin{eqnarray}
\mathrm{\mathbf{E}} & = & \frac{E_{0}}{2}\left(\left[\begin{array}{c}
0\\
1\\
0
\end{array}\right]e^{i\mathbf{k}_{1}\cdot\mathbf{r}}+\left[\begin{array}{c}
\mathrm{cos}\psi\mathrm{sin}\xi\\
\mathrm{cos}\xi\\
\mathrm{sin}\psi\mathrm{sin}\xi
\end{array}\right]e^{i\mathbf{k}_{2}\cdot\mathbf{r}}\right),\label{eq:PWI}
\end{eqnarray}
where $\xi$ is the deviation angle between electric field polarization
of the two plane waves.
The time averaged optical force exerted on an isotropic dipolar particle by the
above illumination can be derived as:
\begin{eqnarray}
\overline{\mathrm{F}}_{x} & = & -\frac{3}{2}\left[\mathrm{Re}\left(\bar{\alpha}_{e}\right)+\mathrm{Re}\left(\bar{\alpha}_{m}\right)\mathrm{cos}2\psi-\mathrm{Re}\left(\bar{\alpha}_{e}\bar{\alpha}_{m}^{*}\right)\right]\nonumber \\
&  & \mathrm{sin}\psi\mathrm{cos}\xi\mathrm{sin}2\delta,\nonumber \\
\overline{\mathrm{F}}_{y} & = & \frac{3}{2}\left[\mathrm{Re}\left(e^{-2i\delta}\bar{\alpha}_{e}\bar{\alpha}_{m}^{*}\right)\right]\mathrm{sin}\psi\mathrm{sin}\xi\mathrm{cos}\psi,\nonumber \\
\overline{\mathrm{F}}_{z} & = & \frac{3}{2}\mathrm{cos}\psi\left\{ \left(1+\mathrm{cos}2\psi\mathrm{cos}\xi\mathrm{cos}2\delta\right)\mathrm{Im}(\bar{\alpha}_{m})\right.\nonumber \\
 &  & \left.+\left[\mathrm{Im}(\bar{\alpha}_{e})-\mathrm{Re}(\bar{\alpha}_{e}\bar{\alpha}_{m}^{*})\right]\left(1+\mathrm{cos}\xi\mathrm{cos}2\delta\right)\right\}. \label{eq:a11}
\end{eqnarray}

Intuitively, one might expect to have an exerted optical force only
in the direction of the overall linear momentum, i.e. $\mathrm{\mathbf{k}}_{1}+\mathrm{\mathbf{k}}_{2}=2k\mathrm{cos}\psi\mathbf{e}_{z}$.
However, according to Eq.~(\ref{eq:a11}), lateral forces (in both
directions of $x$ and $y$) can be experienced by a fully symmetric (isotropic)
particle for certain angles of $\psi$ and $\xi$.
These peculiar lateral forces can be elaborated on the basis of the
symmetry breaking mediated by the illuminating wave rather than the
particle. The variation of lateral optical forces exerted on the dielectric
sphere for parameters $\xi$, $\psi=75^{\circ}$, $x_0 = a/2$ and
 $\psi=75^{\circ},$ $\xi=45^{\circ}$ are illustrated in Fig.~\ref{fig:angle_variation} (a)-(b), respectively.

\section{Conclusion}

In conclusion, we investigated the optical force exerted on an isotropic
particle by two plane waves and demonstrated theoretically that
pushing-pulling forces for either TE or TM illuminations is possible.
Our method, based on the theoretical calculations of multipolar forces,
revealed the contribution of each electric and magnetic moment up
to quadrupole terms (including their interferences) to the optical
force. Additionally, this approach elaborates the optical force in
a closed form due to the electrodynamical formalism of all
influential parameters i.e. the designated angles, polarizabilities
and amplitudes for either TE or TM illuminations. According to this
formalism, we also showed the existence of lateral forces, in both
$x$ and $y$ directions, for certain angles of $\psi$, $\xi$
and an interval of deviation of the object from the center $x_0$. Our approach and findings can be employed in the optical
manipulations and sorting of micro/nano particles with different illuminations.

\section{Acknowledgments} R.A. would like to acknowledge financial support from the Max Planck Society. A.R. acknowledges support from the Karlsruhe School of Optics \& Photonics (KSOP).
\appendix
\begin{widetext}

\section{Optical force: TE illumination}

\subsection{Useful expressions}

\subsubsection{Optical force}

The total optical force exerted on a small particle up to quadrupole
moments (neglecting higher order terms) reads as~\cite{chen2011optical}:

\begin{eqnarray}
\left\langle F_{i}\right\rangle  & = & \frac{1}{2}\textrm{Re}\left[\underset{j}{\sum}p_{j}\nabla_{i}E_{j}^{*}\right]+\frac{1}{2}\textrm{Re}\left[\underset{j}{\sum}m_{j}\nabla_{i}B_{j}^{*}\right]-\frac{k^{4}}{12\pi\varepsilon_{0}c}\textrm{Re}\left[\underset{j,k}{\sum}\epsilon_{ijk}p_{j}m_{k}^{*}\right]\nonumber \\
 & - & \frac{k^{5}}{120\pi\varepsilon_{0}}\textrm{Im}\left[\underset{j}{\sum}\left(Q^{e}\right)_{ij}p_{j}^{*}\right]+\frac{1}{\text{12}}\textrm{Re}\left[\left(Q^{e}\right)_{jk}\nabla_{i}\nabla_{k}E_{j}^{*}\right]+\frac{k^{5}}{120\pi\varepsilon_{0}c^{2}}\textrm{Im}\left[\left(Q^{m}\right)_{ij}m_{j}^{*}\right]\nonumber \\
 & + & \frac{1}{12}\mathrm{Re}\left[\underset{k}{\sum}\left(Q^{m}\right)_{jk}\nabla_{i}\nabla_{k}B_{j}^{*}\right]-\frac{k^{6}}{9\times240\pi\varepsilon_{0}c}\mathrm{Re}\left[\underset{l,j,k}{\sum}\varepsilon_{ijk}(Q^{e})_{lj}\left(Q^{m}\right)_{lk}^{*}\right].\label{eq:a6}
\end{eqnarray}

In the following sections, we will use Eq.~\ref{eq:a6}, to derive
analytical expressions for the exerted optical force on a small particle
up to quadrupolar moments.

\subsubsection{Electric and magnetic fields and and their derivatives}

Here, we derive the whole required electromagnetic fields and their
derivatives which are necessary to derive the exerted optical forces
on small particles by TE and TM illuminations. The TE illumination
is defined as $\mathbf{E}=E_{0}\left(e^{i\mathbf{k}_{1}\cdot\mathbf{r}}+e^{i\mathbf{k}_{2}\cdot\mathbf{r}}\right)\mathbf{e}_{y}/2,$
where $\mathbf{k}_{1}$ and $\mathbf{k}_{2}$ are wave vectors of
the two plane waves and are defined as:

\begin{equation}
\mathbf{k}_{1}=k[\mathrm{sin}\psi,0,\mathrm{cos\psi}]^{T},\,\,\,\,\,\,\,\mathbf{k}_{2}=k[-\mathrm{sin\psi},0,\mathrm{cos\psi}]^{T}.
\end{equation}

The field at any arbitrary point $\mathbf{r}_{0}(x_{0},y_{0},z_{0})$
can also be written as:

\begin{eqnarray}
\mathbf{E}\mid_{\mathbf{r}=\mathbf{r}_{0}} & = & E_{0}\mathrm{cos}\delta e^{ik\mathrm{cos}\psi z_{0}}\mathbf{e}_{y},\label{eq:EField}
\end{eqnarray}

\noindent where $\delta=k\mathrm{sin\psi}x_{0}$. The corresponding
magnetic field is calculated as:

\begin{eqnarray}
\mathbf{B} & = & B_{x}\mathbf{e}_{x}+B_{z}\mathbf{e}_{z}=\frac{1}{\omega}\left[\left(\mathbf{k}_{1}\times\mathbf{E}_{1}\right)+\left(\mathbf{k}_{2}\times\mathbf{E}_{2}\right)\right],\nonumber \\
 & = & \frac{k}{\omega}\mathrm{cos}\textrm{\ensuremath{\psi}}\mathbf{e}_{z}\times\left(\mathbf{E_{\mathrm{1}}}+\mathbf{E}_{2}\right)+\frac{k}{\omega}\mathrm{sin}\textrm{\ensuremath{\psi}}\mathbf{e}_{x}\times\left(\mathbf{E}_{1}-\mathbf{E}_{2}\right),
\end{eqnarray}

\noindent where $\mathbf{E}_{i=1,2}=\frac{E_{0}}{2}e^{i\mathbf{k}_{i}\cdot\mathbf{r}}$$\mathbf{e}_{y}$.
Therefore, the magnetic field at any arbitrary point $\mathbf{r}_{0}(x_{0},y_{0},z_{0})$
can be written as the following:

\begin{eqnarray}
B_{z}\mid_{\mathbf{r}=\mathbf{r}_{0}} & = & iE_{0}\frac{k}{\omega}\mathrm{sin}\psi\mathrm{sin}\delta e^{ik\mathrm{cos}\psi z_{0}},\nonumber \\
B_{x}\mid_{\mathbf{r}=\mathbf{r}_{0}} & = & -E_{0}\frac{k}{\omega}\mathrm{cos}\psi\mathrm{cos}\delta e^{ik\mathrm{cos}\psi z_{0}}.\label{eq:BField}
\end{eqnarray}

Some useful expressions for the derivatives of the electric and magnetic
fields can be calculated as follows:

\begin{eqnarray}
\frac{\partial E_{y}}{\partial x}\mid_{\mathbf{r}=\mathbf{r}_{0}} & = & -E_{0}k\mathrm{sin\psi}\mathrm{sin}\delta e^{ik\mathrm{cos}\psi z_{0}},\nonumber \\
\frac{\partial E_{y}}{\partial x}\mid_{\mathbf{r}=\mathbf{r}_{0}} & = & 0,\nonumber \\
\frac{\partial E_{y}}{\partial x}\mid_{\mathbf{r}=\mathbf{r}_{0}} & = & iE_{0}\mathrm{k\mathrm{cos}\psi cos}\delta e^{ik\mathrm{cos}\psi z_{0}},\label{eq:Derv_E}
\end{eqnarray}

\begin{eqnarray}
\frac{\partial H_{z}}{\partial x}\mid_{\mathbf{r}=\mathbf{r}_{0}} & = & iE_{0}\frac{k^{2}}{\omega\mu}\mathrm{sin^{2}\psi cos\delta}e^{ik\mathrm{cos}\psi z_{0}},\nonumber \\
\frac{\partial H_{x}}{\partial z}\mid_{\mathbf{r}=\mathbf{r}_{0}} & = & -iE_{0}\frac{k^{2}}{\omega\mu}\mathrm{cos^{2}\psi cos\delta e^{ikcos\psi z_{0}}},\nonumber \\
\frac{\partial H_{x}}{\partial x}\mid_{\mathbf{r}=\mathbf{r}_{0}} & = & E_{0}\frac{k^{2}}{\omega\mu}\mathrm{sin\psi}\mathrm{cos}\psi\mathrm{sin}\delta e^{ik\mathrm{cos}\psi z_{0}},\nonumber \\
\frac{\partial H_{z}}{\partial z}\mid_{\mathbf{r}=\mathbf{r}_{0}} & = & -E_{0}\frac{k^{2}}{\omega\mu}\mathrm{sin\psi}\mathrm{cos}\psi\mathrm{sin}\delta e^{ik\mathrm{cos}\psi z_{0}},\label{eq:Derv_H}
\end{eqnarray}

\begin{eqnarray}
\nabla_{z}\nabla_{x}E_{y}^{*}\mid_{\mathbf{r}=\mathbf{r}_{0}} & = & iE_{0}^{*}k^{2}\mathrm{cos}\psi\mathrm{sin}\psi\mathrm{sin}\delta e^{-ik\mathrm{cos}\psi z_{0}},\nonumber \\
\nabla_{z}\nabla_{z}E_{y}^{*}\mid_{\mathbf{r}=\mathbf{r}_{0}} & = & -E_{0}^{*}k^{2}\mathrm{cos^{2}}\psi\mathrm{cos}\delta e^{-ik\mathrm{cos}\psi z_{0}},\nonumber \\
\nabla_{x}\nabla_{x}E_{y}^{*}\mid_{\mathbf{r}=\mathbf{r}_{0}} & = & -E_{0}^{*}k^{2}\mathrm{cos}\delta\mathrm{sin}^{2}\psi e^{-ik\mathrm{cos}\psi z_{0}},\nonumber \\
\nabla_{x}\nabla_{z}E_{y}^{*}\mid_{\mathbf{r}=\mathbf{r}_{0}} & = & iE_{0}^{*}k^{2}\mathrm{sin}\psi\mathrm{cos}\psi\mathrm{sin}\delta e^{-ik\mathrm{cos}\psi z_{0}},\label{eq:Derv_Derv_E}
\end{eqnarray}

\begin{eqnarray}
\nabla_{z}\nabla_{z}B_{x}^{*}\mid_{\mathbf{r}=\mathbf{r}_{0}} & = & E_{0}^{*}\frac{k^{3}}{\omega}\mathrm{cos^{3}\psi cos\delta e^{-ikcos\psi z_{0}}},\nonumber \\
\nabla_{z}\nabla_{x}B_{z}^{*}\mid_{\mathbf{r}=\mathbf{r}_{0}} & = & -E_{0}^{*}\frac{k^{3}}{\omega}\mathrm{sin^{2}\psi cos\psi cos\delta}e^{-ik\mathrm{cos}\psi z_{0}},\nonumber \\
\nabla_{z}\nabla_{x}B_{x}^{*}\mid_{\mathbf{r}=\mathbf{r}_{0}} & = & -iE_{0}^{*}\frac{k^{3}}{\omega}\mathrm{sin\psi}\mathrm{cos^{2}\psi sin\delta e^{-ikcos\psi z_{0}}},\nonumber \\
\nabla_{x}\nabla_{x}B_{x}^{*}\mid_{\mathbf{r}=\mathbf{r}_{0}} & = & E_{0}^{*}\frac{k^{3}}{\omega}\mathrm{sin^{2}}\psi\mathrm{cos\psi cos\delta e^{-i\mathit{k}cos\psi z_{0}}},\nonumber \\
\nabla_{x}\nabla_{x}B_{z}^{*}\mid_{\mathbf{r}=\mathbf{r}_{0}} & = & iE_{0}^{*}\frac{k^{3}}{\omega}\mathrm{sin^{3}}\psi\mathrm{sin\delta e^{-ikcos\psi z_{0}}},\nonumber \\
\nabla_{z}\nabla_{z}B_{z}^{*}\mid_{\mathbf{r}=\mathbf{r}_{0}} & = & iE_{0}^{*}\frac{k^{3}}{\omega}\mathrm{cos^{2}\psi sin\psi sin\delta}e^{-ik\mathrm{cos}\psi z_{0}}.\label{eq:Derv_B}
\end{eqnarray}

\subsubsection{Induced multipole moments}

Electric and magnetic dipole moments for an isotropic object are defined
as $\mathbf{p}=\varepsilon_{0}\alpha_{e}\mathbf{E}$ and $\mathbf{m}=\alpha_{m}\mathbf{H}$,
where $\mathbf{p}$ and $\mathbf{m}$ are the electric and magnetic
dipole moments, respectively. $\alpha_{e}$ and $\alpha_{m}$ denote
the scalar electric and magnetic polarizabilities, respectively. Using
the electric and magnetic fields in Eqs.\textasciitilde{}\ref{eq:EField}-\ref{eq:BField},
the induced moments for an isotropic particle reads as:

\begin{eqnarray}
\mathbf{p} & = & \varepsilon_{0}\alpha_{e}\mathbf{E}=\varepsilon_{0}\alpha_{e}E_{0}\mathrm{cos}\delta e^{\mathrm{i}k\mathrm{cos}\psi z_{0}}\mathbf{e}_{y},\label{eq:ED}
\end{eqnarray}

\begin{eqnarray}
\mathbf{m} & = & \alpha_{m}\mathbf{H}=\alpha_{m}E_{0}\frac{k}{\mu\omega}\left(\mathrm{i}\mathrm{sin}\psi\mathrm{sin}\delta e^{\mathrm{i}k\mathrm{cos}\psi z_{0}}\mathbf{e}_{x}-\mathrm{cos}\psi\mathrm{cos}\delta e^{\mathrm{i}k\mathrm{cos}\psi z_{0}}\mathbf{e}_{z}\right).\label{eq:MD}
\end{eqnarray}

The electric quadrupole moment ($\mathbf{Q}^{e}$) induced in an isotropic
particle is defined as:

\begin{equation}
\mathbf{Q}^{e}=\left[\begin{array}{ccc}
Q_{xx}^{e} & Q_{xy}^{e} & Q_{xz}^{e}\\
Q_{xy}^{e} & Q_{yy}^{e} & Q_{yz}^{e}\\
Q_{xz}^{e} & Q_{yz}^{e} & Q_{zz}^{e}
\end{array}\right],
\end{equation}

\noindent where the elements of the matrix are defined as $Q_{ij}^{e}=\varepsilon_{0}\alpha_{Q^{e}}\left(\partial_{i}E_{j}+\partial_{j}E_{i}\right)/2$
, $i,j=1,2,3$ for $x$, $y$ and $z$ \cite{Alu:2009,bernal2014underpinning}.
$\mathbf{Q}^{e}$ is a second rank tensor and a traceless matrix,
i.e. $Q_{xx}^{e}+Q_{yy}^{e}+Q_{zz}^{e}=0$. Using Eq.~\ref{eq:Derv_E},
we can calculate all the components of the electric quadrupole moments:

\begin{eqnarray}
Q_{xx}^{e} & = & \varepsilon_{0}\alpha_{Q^{e}}\frac{\partial_{x}E_{x}+\partial_{x}E_{x}}{2}=0,\nonumber \\
Q_{yy}^{e} & = & \varepsilon_{0}\alpha_{Q^{e}}\frac{\partial_{y}E_{y}+\partial_{y}E_{y}}{2}=0,\nonumber \\
Q_{zz}^{e} & = & \varepsilon_{0}\alpha_{Q^{e}}\frac{\partial_{z}E_{z}+\partial_{z}E_{z}}{2}=0,\nonumber \\
Q_{zx}^{e} & = & \varepsilon_{0}\alpha_{Q^{e}}\frac{\partial_{z}E_{x}+\partial_{x}E_{z}}{2}=0,\nonumber \\
Q_{yz}^{e} & = & \varepsilon_{0}\alpha_{Q^{e}}\frac{\partial_{y}E_{z}+\partial_{z}E_{y}}{2}=\varepsilon_{0}\alpha_{Q^{e}}\frac{\partial_{z}E_{y}}{2},\nonumber \\
 & = & \frac{ikE_{0}}{2}\varepsilon_{0}\alpha_{Q^{e}}\mathrm{cos\psi\mathrm{cos}\delta}e^{ik\mathrm{cos}\psi z_{0}},\nonumber \\
Q_{yx}^{e} & = & \varepsilon_{0}\alpha_{Q^{e}}\frac{\partial_{y}E_{x}+\partial_{x}E_{y}}{2}=\varepsilon_{0}\alpha_{Q^{e}}\frac{\partial_{x}E_{y}}{2},\nonumber \\
 & = & -\frac{kE_{0}}{2}\varepsilon_{0}\alpha_{Q^{e}}\mathrm{sin\psi}\mathrm{sin}\delta e^{ik\mathrm{cos}\psi z_{0}}.\label{eq:Qe}
\end{eqnarray}

Similarly, the magnetic quadrupole moment ($\mathbf{Q}^{m}$) for
an isotropic particle is defined as :

\begin{equation}
Q^{m}=\left[\begin{array}{ccc}
Q_{xx}^{m} & Q_{xy}^{m} & Q_{xz}^{m}\\
Q_{xy}^{m} & Q_{yy}^{m} & Q_{yz}^{m}\\
Q_{xz}^{m} & Q_{yz}^{m} & Q_{zz}^{m}
\end{array}\right],
\end{equation}

\noindent where the elements of the matrix are defined as $Q_{ij}^{m}=\alpha_{Q^{m}}\left(\partial_{i}H_{j}+\partial_{j}H_{i}\right)/2$,
and $i,j=1,2,3$ for $x$, $y$ and $z$ \cite{Alu:2009,bernal2014underpinning}.
$\mathbf{Q}^{m}$ is a second rank tensor and a traceless matrix,
i.e. $Q_{xx}^{m}+Q_{yy}^{m}+Q_{zz}^{m}=0$. Using the magnetic field
in Eq.~\ref{eq:BField} and its derivative Eq.~\ref{eq:Derv_H},
we can calculate all components of the magnetic quadrupole moments:

\begin{eqnarray}
Q_{yy}^{m} & = & \alpha_{Q^{m}}\frac{\partial_{y}H_{y}+\partial_{y}H_{y}}{2}=0,\nonumber \\
Q_{yx}^{m} & = & \alpha_{Q^{m}}\frac{\partial_{y}H_{x}+\partial_{x}H_{y}}{2}=0,\nonumber \\
Q_{xx}^{m} & = & \alpha_{Q^{m}}\frac{\partial_{x}H_{x}+\partial_{x}H_{x}}{2}=\alpha_{Q^{m}}\partial_{x}H_{x},\nonumber \\
 & = & E_{0}\frac{k^{2}}{\omega\mu}\alpha_{Q^{m}}\mathrm{sin\psi cos}\psi\mathrm{sin}\delta e^{\mathrm{i}k\mathrm{cos\psi}z_{0}},\nonumber \\
Q_{zz}^{m} & = & -Q_{xx}^{m},\nonumber \\
Q_{yz}^{m} & = & \alpha_{Q^{m}}\frac{\partial_{y}H_{z}+\partial_{z}H_{y}}{2}=0,\nonumber \\
Q_{xz}^{m} & = & \alpha_{Q^{m}}\frac{\partial_{x}H_{z}+\partial_{z}H_{x}}{2},\nonumber \\
 & = & \mathrm{i}\frac{E_{0}}{2}\frac{k^{2}}{\omega\mu}\alpha_{Q^{m}}\left(\mathrm{sin^{2}\psi-cos^{2}\psi}\right)\mathrm{cos}\delta e^{\mathrm{i}k\mathrm{cos\psi}z}.\label{eq:Qm}
\end{eqnarray}

In the following section, we calculate the components of the optical
force by using the induced multipole moments, the electric and magnetic
fields and their derivatives.

\subsection{$\mathbf{F}_{p}$ contribution }

According to Eq.~\ref{eq:a6}, the electric dipole contribution reads
as:

\begin{equation}
F_{i(p)}=\frac{1}{2}\mathrm{Re}\left(\underset{j}{\sum}p_{j}\nabla_{i}E_{j}^{*}\right),\label{eq:5}
\end{equation}

\noindent $i,j=1,2,3$ for $x$, $y$ and $z$. Using the electric
field in Eq.~\ref{eq:EField}, and the definition of the electric
dipole moment $\mathbf{p}=\epsilon_{0}\alpha_{e}\mathbf{E}$, it can
be easily seen that the $y$ component of $\mathbf{F}_{p}$ is zero.

Using Eq.~\ref{eq:5}, the $x$ component of $\mathbf{F}_{p}$ reads
as:

\begin{eqnarray}
F_{x(p)} & = & \frac{1}{2}\mathrm{Re}\left(p_{x}\frac{\partial}{\partial x}E_{x}^{*}+p_{y}\frac{\partial}{\partial x}E_{y}^{*}+p_{z}\frac{\partial}{\partial x}E_{z}^{*}\right)=\frac{1}{2}\mathrm{Re}\left(p_{y}\frac{\partial}{\partial x}E_{y}^{*}\right).\label{eq:Fpx}
\end{eqnarray}

Now, by substituting Eq.~\ref{eq:ED} and Eq.~\ref{eq:Derv_E} into
Eq.~\ref{eq:Fpx}, we obtain:

\begin{eqnarray*}
F_{x(p)} & = & \frac{1}{2}\mathrm{Re}\left[\varepsilon_{0}\left|E_{0}^{2}\right|\alpha_{e}\mathrm{cos}\delta\mathrm{sin}\delta\left(-k\mathrm{sin}\psi\right)\right],\\
 & = & -3\left(\frac{k^{3}}{6\pi}\right)F^{\mathrm{norm}}\mathrm{sin\psi}\textrm{\ensuremath{\mathrm{cos}\delta\mathrm{sin}\delta}}\mathrm{Re}\left(\alpha_{e}\right),\\
 & = & -3\left(\frac{k^{3}}{12\pi}\right)F^{\mathrm{norm}}\mathrm{sin\psi}\textrm{\ensuremath{\mathrm{sin2}\delta}}\mathrm{Re}\left(\alpha_{e}\right).
\end{eqnarray*}

Finally, by using the definition of the normalized force, i.e. $\bar{\mathbf{F}}=\mathbf{F}/\mathrm{F}^{\mathrm{norm}}$
and the normalized electric polarizability, i.e. $\bar{\alpha}_{e}=\alpha_{e}/\alpha_{d}$,
we obtain

\begin{equation}
\boxed{\bar{F}_{x(p)}=-\frac{3}{2}\mathrm{sin\psi}\textrm{\ensuremath{\mathrm{sin2}\delta}}\mathrm{Re}(\bar{\alpha}_{e}),}
\end{equation}

\noindent where $\alpha_{d}=6\pi/k^{3}$, and $\mathrm{F}^{\mathrm{norm}}=\frac{I_{0}}{c}\frac{\lambda^{2}}{2\pi}$.
This expression is documented in Eq.~7 of the main manuscript.

Similarly, the $z$ component of $\mathbf{F}_{p}$ read as:

\begin{eqnarray}
F_{z(p)} & = & \frac{1}{2}\mathrm{Re}\left(p_{x}\frac{\partial}{\partial z}E_{x}^{*}+p_{y}\frac{\partial}{\partial z}E_{y}^{*}+p_{z}\frac{\partial}{\partial z}E_{z}^{*}\right)=\frac{1}{2}\mathrm{Re}\left(p_{y}\frac{\partial}{\partial z}E_{y}^{*}\right).\label{eq:Fpz}
\end{eqnarray}

Now, by substituting Eq.~\ref{eq:ED} and Eq.~\ref{eq:Derv_E} into
Eq.~\ref{eq:Fpz}, we obtain:

\begin{eqnarray*}
F_{z(p)} & = & \frac{1}{2}\mathrm{Re}\left[\varepsilon_{0}\left|E_{0}^{2}\right|\alpha_{e}\mathrm{cos^{2}}\left(\delta\right)\left(-\mathrm{i}k\mathrm{cos}\psi\right)\right],\\
 & = & 3\left(\frac{k^{3}}{6\pi}\right)F^{\mathrm{norm}}\mathrm{cos}\psi\textrm{\ensuremath{\mathrm{cos^{2}}\delta\mathrm{Im}}}\left(\alpha_{e}\right).
\end{eqnarray*}

Then, the normalized contribution is derived as:

\begin{equation}
\boxed{\bar{F}_{z(p)}=3\mathrm{cos}\psi\mathrm{cos^{2}}\delta\textrm{\ensuremath{\mathrm{Im}}}\left(\bar{\alpha}_{e}\right),}
\end{equation}

This expression is documented in Eq.~7 of the main manuscript.

\subsection{$\mathbf{F}_{m}$ contribution }

According to Eq.~\ref{eq:a6}, the magnetic dipole contribution reads
as:

\begin{equation}
F_{i(m)}=\frac{1}{2}\mathrm{Re}\left(\underset{j}{\sum}m_{j}\nabla_{i}B_{j}^{*}\right),\label{eq:2}
\end{equation}

$i,j=1,2,3$ for $x$, $y$ and $z$. Using the magnetic field, i.e.
Eq.~\ref{eq:BField}, and the definition of the magnetic dipole moment
$\mathbf{m}=\alpha_{m}\mathbf{H}$, it can be easily seen that the
$y$ component of the $\mathbf{F}_{m}$ is zero.

Using Eq.~\ref{eq:2}, the $x$ component of $\mathbf{F}_{m}$ read
as

\begin{eqnarray}
F_{x(m)} & = & \frac{1}{2}\mathrm{Re}\left(m_{x}\frac{\partial}{\partial x}B_{x}^{*}+m_{y}\frac{\partial}{\partial x}B_{y}^{*}+m_{z}\frac{\partial}{\partial x}B_{z}^{*}\right),\nonumber \\
 & = & \frac{1}{2}\mathrm{Re}\left(m_{x}\frac{\partial}{\partial x}B_{x}^{*}+m_{z}\frac{\partial}{\partial x}B_{z}^{*}\right).\label{eq:Fmx}
\end{eqnarray}

Now, by substituting Eq.~\ref{eq:MD} and Eq.~\ref{eq:Derv_H} into
Eq.~\ref{eq:Fmx}, we obtain:

\begin{eqnarray}
F_{x(m)} & = & \frac{1}{2}\mathrm{Re}\left[\alpha_{m}\left(\mathrm{\mathit{k}}\mathrm{sin\psi}\right)\frac{1}{\mu}(\frac{k}{\omega})^{2}\left(\mathrm{sin^{2}}\psi-\mathrm{cos^{2}}\psi\right)\mathrm{cos}\delta\mathrm{sin}\delta\left|E_{0}\right|^{2}\right],\nonumber \\
 & = & 3\left(\frac{k^{3}}{6\pi}\right)F^{\mathrm{norm}}\mathrm{sin}\psi\mathrm{cos}\delta\mathrm{sin}\delta\left(\mathrm{sin^{2}}\psi-\mathrm{cos^{2}}\psi\right)\mathrm{Re}\left(\alpha_{m}\right),\nonumber \\
 & = & -3\left(\frac{k^{3}}{6\pi}\right)F^{\mathrm{norm}}\mathrm{sin}\psi\mathrm{cos}\delta\mathrm{sin}\delta\mathrm{cos2}\psi\mathrm{Re}\left(\alpha_{m}\right),\nonumber \\
 & = & -\frac{3}{2}\left(\frac{k^{3}}{6\pi}\right)F^{\mathrm{norm}}\mathrm{sin}\psi\mathrm{sin}2\delta\mathrm{cos}2\psi\mathrm{Re}\left(\alpha_{m}\right).
\end{eqnarray}

Finally, by using the definition of the normalized force, i.e. $\bar{\mathbf{F}}=\mathbf{F}/\mathrm{F}^{\mathrm{norm}}$
and the normalized magnetic polarizability, i.e. $\bar{\alpha}_{m}=\alpha_{m}/\alpha_{d}$,
we obtain:

\begin{equation}
\boxed{\bar{F}_{x(m)}=-\frac{3}{2}\mathrm{sin}\psi\mathrm{sin}2\delta\mathrm{cos}2\psi\mathrm{Re}\left(\bar{\alpha}_{m}\right).}
\end{equation}

This expression is documented in Eq.~7 of the main manuscript.

Similarly, the $z$ component of $\mathbf{F}_{m}$ reads as:

\begin{eqnarray}
F_{z(m)} & = & \frac{1}{2}\mathrm{Re}\left(m_{x}\frac{\partial}{\partial z}B_{x}^{*}+m_{y}\frac{\partial}{\partial z}B_{y}^{*}+m_{z}\frac{\partial}{\partial z}B_{z}^{*}\right),\nonumber \\
 & = & \frac{1}{2}\mathrm{Re}\left(m_{x}\frac{\partial}{\partial z}B_{x}^{*}+m_{z}\frac{\partial}{\partial z}B_{z}^{*}\right).\label{eq:Fmz}
\end{eqnarray}

Now, by substituting Eq.~\ref{eq:MD} and Eq.~\ref{eq:Derv_H} into
Eq.~\ref{eq:Fmz}, we obtain

\begin{eqnarray}
F_{z(m)} & = & \frac{1}{2}\mathrm{Re}\left(-\mathrm{i}\varepsilon_{0}k\mathrm{cos^{3}}\psi\mathrm{cos^{2}}\delta\alpha_{m}\left|E_{0}\right|^{2}\right)\nonumber \\
 &  & +\frac{1}{2}\mathrm{Re}\left(-\mathrm{i}\varepsilon_{0}k\mathrm{cos}\psi\mathrm{sin^{2}}\psi\mathrm{sin^{2}}\delta\alpha_{m}\left|E_{0}\right|^{2}\right),\nonumber \\
 & = & 3\left(\frac{k^{3}}{6\pi}\right)F^{\mathrm{norm}}\mathrm{cos}\psi\left(\mathrm{cos^{2}}\psi\mathrm{cos^{2}}\delta+\mathrm{sin^{2}}\psi\mathrm{sin^{2}}\delta\right)\mathrm{Im}\left(\alpha_{m}\right),
\end{eqnarray}

Finally, the normalized contribution is derived as:

\begin{equation}
\boxed{\bar{F}_{z(m)}=3\mathrm{cos}\psi\left(\mathrm{cos^{2}}\psi\mathrm{cos^{2}}\delta+\mathrm{sin^{2}}\psi\mathrm{sin^{2}}\delta\right)\mathrm{Im}\left(\bar{\alpha}_{m}\right),}
\end{equation}

this expression is documented in Eq.~7 of the main manuscript.

\subsection{$\mathbf{F}_{pm}$ contribution }

According to Eq.~\ref{eq:a6}, the interference dipolar term reads
as:

\begin{equation}
F_{i(pm)}=-\frac{k^{4}}{12\pi\varepsilon_{0}c}\mathrm{Re}\left(\underset{j,k}{\sum}\epsilon_{ijk}p_{j}m_{k}^{*}\right),\label{eq:7}
\end{equation}

\noindent where $\epsilon_{ijk}$ is the \textit{Levi-Civita} symbol
and is defined as:

\[
\epsilon_{ijk}=\begin{cases}
+1 & if\:(i,j,k)\;is\;(1,2,3),(2,3,1),(3,1,2)\\
-1 & if\:(i,j,k)\;is\;(3,2,1),(1,3,2),(2,1,3)\\
0 & if\;i=j\;or\;j=k\;or\;i=k\,\,\,\,\,.
\end{cases}
\]

Using Eq.~\ref{eq:7}, the $x$ component of $\mathbf{F}_{pm}$ read
as

\begin{eqnarray}
F_{x(pm)} & = & \frac{-k^{4}}{12\pi\varepsilon_{0}c}\mathrm{Re}\left(\epsilon_{123}p_{y}m_{z}^{*}+\epsilon_{132}p_{z}m_{y}^{*}\right).\label{eq:Fpmx}
\end{eqnarray}

Now, by substituting Eq.~\ref{eq:ED} and Eq.~\ref{eq:MD} into
Eq.~\ref{eq:Fpmx}, we obtain:

\begin{eqnarray}
F_{x(pm)} & = & \frac{-k^{4}}{12\pi\varepsilon_{0}c}\mathrm{Re}\left[\left(\varepsilon_{0}\alpha_{e}E_{y}\right)\left(\alpha_{m}\frac{B_{z}}{\mu}\right)^{*}\right],\nonumber \\
 & = & \frac{-k^{4}}{12\pi}\mathrm{Re}\left[-i\varepsilon_{0}\alpha_{e}\alpha_{m}^{*}\mathrm{sin}\psi\mathrm{cos}\delta\mathrm{sin}\delta\left|E_{0}\right|^{2}\right],\nonumber \\
 & = & -\frac{3}{2}F^{\mathrm{norm}}\mathrm{sin}\psi\mathrm{sin}2\delta\mathrm{Im}\left[\left(\frac{k^{3}}{6\pi}\right)^{2}\left(\alpha_{e}\alpha_{m}^{*}\right)\right].
\end{eqnarray}

Finally, by using the definition of the normalized force and the normalized
polarizabilities, we obtain:

\begin{equation}
\boxed{\bar{F}_{x(pm)}=-\frac{3}{2}\mathrm{sin}\psi\mathrm{sin}2\delta\mathrm{Im}\left[\frac{k^{3}}{6\pi}\left(\bar{\alpha}_{e}\bar{\alpha}_{m}^{*}\right)\right].}
\end{equation}

This expression is documented in Eq.~7 of the main manuscript.

Using Eq.~\ref{eq:7}, the $z$ component of $\mathbf{F}_{pm}$ read
as

\begin{eqnarray}
F_{z(pm)} & = & \frac{-k^{4}}{12\pi\varepsilon_{0}c}\mathrm{Re}\left(\epsilon_{312}p_{x}m_{y}^{*}+\epsilon_{321}p_{y}m_{x}^{*}\right),\label{eq:Fpmz}
\end{eqnarray}

Now, by substituting Eq.~\ref{eq:ED} and Eq.~\ref{eq:MD} into
Eq.~\ref{eq:Fpmx}, we obtain

\begin{eqnarray*}
F_{z(pm)} & = & \frac{-k^{4}}{12\pi\varepsilon_{0}c}\mathrm{Re}\left[\left(-1\right)\left(\varepsilon_{0}\alpha_{e}E_{y}\right)\left(\alpha_{m}\frac{B_{x}}{\mu}\right)^{*}\right]\\
 & = & \frac{-k^{4}}{12\pi}\mathrm{Re}\left[\varepsilon_{0}\alpha_{e}\alpha_{m}^{*}\mathrm{cos}\psi\mathrm{cos^{2}}\delta\left|E_{0}\right|^{2}\right]\\
 & = & -3F^{\mathrm{norm}}\mathrm{cos}\psi\mathrm{cos^{2}}\delta\mathrm{Re}\left[\left(\frac{k^{3}}{6\pi}\right)^{2}\left(\alpha_{e}\alpha_{m}^{*}\right)\right]
\end{eqnarray*}

Finally, the normalized contribution derives as:

\begin{equation}
\boxed{\bar{F}_{z(pm)}=-3\mathrm{cos}\psi\mathrm{cos^{2}}\delta\mathrm{Re}\left(\bar{\alpha}_{e}\bar{\alpha}_{m}^{*}\right),}
\end{equation}

this expression is documented in Eq.~7 of the main manuscript.

\subsection{$\mathbf{F}_{pQ^{e}}$ contribution }

According to Eq.~\ref{eq:a6}, the optical force caused by the interference
of the electrical dipole and electric quadrupole reads as:

\begin{equation}
F_{i(pQ^{e})}=-\frac{k^{5}}{120\pi\varepsilon_{0}}\mathrm{Im}\left[\underset{j}{\sum}\left(Q^{e}\right)_{ij}p_{j}^{*}\right].\label{eq:8}
\end{equation}

Using Eq.~\ref{eq:8}, the $x$ component of $\mathbf{F}_{pQ^{e}}$
reads as:

\begin{eqnarray}
F_{x(pQ^{e})} & = & -\frac{k^{5}}{120\pi\varepsilon_{0}}\mathrm{Im}\left(Q_{xy}^{e}p_{y}^{*}\right).\label{eq:FpQe_x}
\end{eqnarray}

Now, by substituting Eq.~\ref{eq:ED} and Eq.~\ref{eq:Qe} into
Eq.~\ref{eq:FpQe_x}, we obtain:

\begin{eqnarray*}
F_{x(pQ^{e})} & = & \frac{k^{5}}{120\pi}\frac{1}{2}\varepsilon_{0}k\left|E_{0}\right|^{2}\mathrm{sin}\psi\mathrm{cos}\delta\mathrm{sin}\delta\mathrm{Im}\left(\alpha_{Q^{e}}\alpha_{e}^{*}\right),\\
 & = & \frac{3}{2}F^{\mathrm{norm}}\mathrm{sin}\psi\mathrm{sin}2\delta\mathrm{Im}\left(\frac{k^{3}}{6\pi}\frac{k^{5}}{120\pi}\alpha_{Q^{e}}\alpha_{e}^{*}\right).
\end{eqnarray*}

Finally, by using the definition of the normalized force, i.e. $\bar{\mathbf{F}}=\mathbf{F}/\mathrm{F}^{\mathrm{norm}}$
and the normalized electric dipolar and quadrupolar polarizabilities,
i.e. $\bar{\alpha}_{e}=\alpha_{e}/\alpha_{d}$, $\bar{\alpha}_{Q^{e}}=\alpha_{Q^{e}}/\alpha_{q}$,
we obtain:
\begin{equation}
\boxed{\bar{F}_{x(pQ^{e})}=\frac{3}{2}\mathrm{sin}\psi\mathrm{sin}\left(2\delta\right)\mathrm{Im}\left(\bar{\alpha}_{Q^{e}}\bar{\alpha}_{e}^{*}\right),}
\end{equation}

\noindent where $\alpha_{q}=120\pi/k^{5}$. This expression is documented
in Eq.~7 of the main manuscript.

Similarly, using Eq.~\ref{eq:8}, the $z$ component of $\mathbf{F}_{pQ^{e}}$
reads as:

\begin{eqnarray}
F_{z(pQ^{e})} & = & -\frac{k^{5}}{120\pi\varepsilon_{0}}\mathrm{Im}\left(Q_{zy}^{e}p_{y}^{*}\right).\label{eq:FpQe_z}
\end{eqnarray}

Now, by substituting Eq.~\ref{eq:ED} and Eq.~\ref{eq:Qe} into
Eq.~\ref{eq:FpQe_x}, we obtain

\begin{eqnarray*}
F_{z(pQ^{e})} & = & -\frac{k^{5}}{120\pi}\frac{1}{2}\varepsilon_{0}k\left|E_{0}\right|^{2}\mathrm{cos}\psi\mathrm{cos^{2}}\delta\mathrm{Im}\left(\mathrm{i}\alpha_{Q^{e}}\alpha_{e}^{*}\right),\\
 & = & -3F^{\mathrm{norm}}\mathrm{cos}\psi\mathrm{cos^{2}}\delta\mathrm{Re}\left(\frac{k^{3}}{6\pi}\frac{k^{5}}{120\pi}\alpha_{Q^{e}}\alpha_{e}^{*}\right).
\end{eqnarray*}

Finally, the normalized contribution is derived as:

\begin{equation}
\boxed{\bar{F}_{z(pQ^{e})}=-3\mathrm{cos}\psi\mathrm{cos^{2}}\delta\mathrm{Re}\left(\bar{\alpha}_{Q^{e}}\bar{\alpha}_{e}^{*}\right).}
\end{equation}

This expression is documented in Eq.~7 of the main manuscript.

\subsection{$\mathbf{F}_{Q^{e}}$ contribution }

According to Eq.~\ref{eq:a6}, the electric quadrupole ($Q^{e}$)
contribution reads as:

\begin{equation}
F_{i(Q^{e})}=\frac{1}{12}\mathrm{Re}\left[\underset{k}{\sum}\left(Q_{e}\right)_{jk}\nabla_{i}\nabla_{k}E_{j}^{*}\right].\label{eq:9}
\end{equation}

Using Eq.~\ref{eq:9}, the $x$ component of $\mathbf{F}_{Q^{e}}$
read as

\begin{eqnarray}
F_{x(Q^{e})} & = & \frac{1}{12}\mathrm{Re}\left[\underset{k}{\sum}\left(Q_{e}\right)_{yk}\nabla_{x}\nabla_{k}E_{y}^{*}\right],\nonumber \\
 & = & \frac{1}{12}\mathrm{Re}\left[\left(Q_{e}\right)_{yx}\nabla_{x}\nabla_{x}E_{y}^{*}+\left(Q_{e}\right)_{yz}\nabla_{x}\nabla_{z}E_{y}^{*}\right],\label{eq:FQe_x}
\end{eqnarray}

Now, by substituting Eq.~\ref{eq:Derv_Derv_E} and Eq.~\ref{eq:Qe}
into Eq.~\ref{eq:FQe_x}, we obtain

\begin{eqnarray*}
F_{x(Q^{e})} & = & \frac{1}{12}\mathrm{Re}\left[-\frac{1}{2}\varepsilon_{0}\alpha_{Q^{e}}E_{0}k\mathrm{sin\psi}\mathrm{sin}\left(\delta\right)\left(-E_{0}^{*}k^{2}\mathrm{cos}\delta\mathrm{sin}^{2}\psi\right)\right],\\
 &  & +\frac{1}{12}\mathrm{Re}\left[\frac{1}{2}i\varepsilon_{0}\alpha_{Q^{e}}E_{0}k\mathrm{cos\psi}\mathrm{cos}\left(\delta\right)\left(iE_{0}^{*}k^{2}\mathrm{cos}\psi\mathrm{sin}\psi\mathrm{sin}\delta\right)\right],\\
 & = & -\frac{5}{2}F^{\mathrm{norm}}\mathrm{sin}\psi\mathrm{sin}2\delta\mathrm{cos2\psi Re}\left(\frac{k^{5}}{120\pi}\alpha_{Q^{e}}\right),
\end{eqnarray*}

Finally, by using the definition normalized force and the normalized
polarizabilities, we obtain:

\[
\boxed{\bar{F}_{x(Q^{e})}=-\frac{5}{2}F_{p}^{\mathrm{norm}}\mathrm{sin}\psi\mathrm{sin}2\delta\mathrm{cos2\psi Re}\left(\bar{\alpha}_{Q^{e}}\right).}
\]

This expression is documented in Eq.~7 of the main manuscript.

Using Eq.~\ref{eq:9}, the $z$ component of $\mathbf{F}_{Q^{e}}$
reads as:

\begin{eqnarray}
F_{z(Q^{e})} & = & \frac{1}{12}\mathrm{Re}\left[\underset{k}{\sum}\left(Q_{e}\right)_{yk}\nabla_{z}\nabla_{k}E_{y}^{*}\right],\nonumber \\
 & = & \frac{1}{12}\mathrm{Re}\left[\left(Q_{e}\right)_{yx}\nabla_{z}\nabla_{x}E_{y}^{*}+\left(Q_{e}\right)_{yz}\nabla_{z}\nabla_{z}E_{y}^{*}\right].\label{eq:FQe_z}
\end{eqnarray}

Now, by substituting Eq.~\ref{eq:Derv_Derv_E} and Eq.~\ref{eq:Qe}
into Eq.~\ref{eq:FQe_z}, we obtain:

\begin{eqnarray*}
F_{z(Q^{e})} & = & \frac{1}{12}\mathrm{Re}\left[-\frac{1}{2}\varepsilon_{0}\alpha_{Q^{e}}E_{0}k\mathrm{sin\psi}\mathrm{sin}\delta\left(iE_{0}^{*}k^{2}\mathrm{cos}\psi\mathrm{sin}\psi\mathrm{sin}\delta\right)\right],\\
 & = & 5F^{\mathrm{norm}}\mathrm{cos}\psi\left(\mathrm{cos}^{2}\psi\mathrm{cos^{2}}\delta+\mathrm{sin^{2}}\psi\mathrm{sin^{2}}\delta\right)\mathrm{Im}\left(\frac{k^{5}}{120\pi}\alpha_{Q^{e}}\right).
\end{eqnarray*}

Finally, the normalized contribution is:

\begin{equation}
\boxed{\bar{F}_{z(Q^{e})}=5\mathrm{cos}\psi\left(\mathrm{cos}^{2}\psi\mathrm{cos^{2}}\delta+\mathrm{sin^{2}}\psi\mathrm{sin^{2}}\delta\right)\mathrm{Im}\left(\bar{\alpha}_{Q^{e}}\right).}
\end{equation}

This expression is documented in Eq.~7 of the main manuscript.

\subsection{$\mathbf{F}_{Q^{m}}$ contribution }

According to Eq.~\ref{eq:a6}, the magnetic quadrupole($Q^{m}$)
contribution is given by:

\begin{equation}
F_{i(Q^{m})}=\frac{1}{12}\mathrm{Re}\left[\underset{k}{\sum}\left(Q^{m}\right)_{jk}\nabla_{i}\nabla_{k}B_{j}^{*}\right].\label{eq:10}
\end{equation}

Using Eq.~\ref{eq:10}, the $x$ component of $\mathbf{F}_{Q^{m}}$
read as

\begin{eqnarray}
F_{x(Q^{m})} & = & \frac{1}{12}\mathrm{Re}\left[\underset{k}{\sum}\left(Q^{m}\right)_{xk}\nabla_{x}\nabla_{k}B_{x}^{*}\right]+\frac{1}{12}\mathrm{Re}\left[\underset{k}{\sum}\left(Q^{m}\right)_{zk}\nabla_{x}\nabla_{k}B_{z}^{*}\right],\label{eq:FQm_x}\\
 & = & \frac{1}{12}\mathrm{Re}\left[\left(Q^{m}\right)_{xx}\nabla_{x}\nabla_{x}B_{x}^{*}+\left(Q^{m}\right)_{xy}\nabla_{x}\nabla_{y}B_{x}^{*}+\left(Q^{m}\right)_{xz}\nabla_{x}\nabla_{z}B_{x}^{*}\right],\nonumber \\
 & + & \frac{1}{12}\mathrm{Re}\left[\left(Q^{m}\right)_{zx}\nabla_{x}\nabla_{x}B_{z}^{*}+\left(Q^{m}\right)_{zy}\nabla_{x}\nabla_{y}B_{z}^{*}+\left(Q^{m}\right)_{zz}\nabla_{x}\nabla_{z}B_{z}^{*}\right],\nonumber \\
 & = & \frac{1}{12}\mathrm{Re}\left[\left(Q^{m}\right)_{xz}\nabla_{x}\nabla_{z}B_{x}^{*}+\left(Q^{m}\right)_{xx}\nabla_{x}\nabla_{x}B_{x}^{*}+\left(Q^{m}\right)_{zx}\nabla_{x}\nabla_{x}B_{z}^{*}+\left(Q^{m}\right)_{zz}\nabla_{x}\nabla_{z}B_{z}^{*}\right].\nonumber
\end{eqnarray}

Now, by substituting Eq.~\ref{eq:Derv_B} and Eq.~\ref{eq:Qm} into
Eq.~\ref{eq:FQm_x}, we obtain:

\begin{eqnarray*}
F_{x(Q^{m})} & = & \frac{1}{12}\mathrm{Re}\left\{ \mathrm{i}\alpha_{Q^{m}}\frac{E_{0}}{2}\left(\frac{k^{2}}{\omega\mu}\left(\mathrm{sin^{2}\psi-cos^{2}\psi}\right)\mathrm{cos}\delta\right)\left[-\mathrm{i}E_{0}^{*}\left(\frac{k^{3}}{\omega}\mathrm{sin\psi}\mathrm{cos^{2}\psi sin\delta}\right)\right]\right\} \\
 & + & \frac{1}{12}\mathrm{Re}\left[\alpha_{Q^{m}}E_{0}\frac{k^{2}}{\omega\mu}\mathrm{sin}\psi\mathrm{cos}\psi\mathrm{sin}\delta E_{0}^{*}\left(\frac{k^{3}}{\omega}\mathrm{sin^{2}}\psi\mathrm{cos\psi cos\delta}\right)\right]\\
 & + & \frac{1}{12}\mathrm{Re}\left\{ \mathrm{i}\alpha_{Q^{m}}\frac{E_{0}}{2}\left(\frac{k^{2}}{\omega\mu}\left(\mathrm{sin^{2}\psi-cos^{2}\psi}\right)\mathrm{cos}\delta\right)\left[\mathrm{i}E_{0}^{*}\left(\frac{k^{3}}{\omega}\mathrm{sin^{3}}\psi\mathrm{sin\delta}\right)\right]\right\} \\
 & + & \frac{1}{12}\mathrm{Re}\left[-\alpha_{Q^{m}}E_{0}\frac{k^{2}}{\omega\mu}\mathrm{sin}\psi\mathrm{cos}\psi\mathrm{sin}\delta\left(-E_{0}^{*}\right)\left(\frac{k^{3}}{\omega}\mathrm{sin^{2}\psi cos\psi cos\delta}\right)\right]\\
 & = & -\frac{5}{2}F_{p}^{\mathrm{norm}}\mathrm{sin\psi\mathrm{sin}2\delta}\left(\mathrm{cos^{2}}2\psi-\mathrm{sin^{2}}2\psi\right)\mathrm{Re}\left(\frac{k^{5}}{120\pi}\alpha_{Q^{m}}\right).
\end{eqnarray*}

Finally, by using the definition of the normalized force and the normalized
magnetic quadrupolar polarizabilities, i.e. $\bar{\alpha}_{Q^{m}}=\alpha_{Q^{m}}/\alpha_{q}$,
we obtain:

\begin{equation}
\boxed{\bar{F}_{x(Q^{m})}=-\frac{5}{2}\mathrm{sin\psi\mathrm{sin}2\delta\left(\mathrm{cos^{2}}2\psi-\mathrm{sin^{2}}2\psi\right)}\mathrm{Re}\left(\bar{\alpha}_{Q^{m}}\right).}
\end{equation}

This expression is documented in Eq.~7 of the main manuscript.

Using Eq.~\ref{eq:10}, the $z$ component of $\mathbf{F}_{Q^{m}}$
reads as:

\begin{eqnarray}
F_{z(Q^{m})} & = & \frac{1}{12}\mathrm{Re}\left[\underset{k}{\sum}\left(Q^{m}\right)_{xk}\nabla_{z}\nabla_{k}B_{x}^{*}\right]+\frac{1}{12}\mathrm{Re}\left[\underset{k}{\sum}\left(Q^{m}\right)_{zk}\nabla_{z}\nabla_{k}B_{z}^{*}\right],\label{eq:FQm_z}\\
 & = & \frac{1}{12}\mathrm{Re}\left[\left(Q^{m}\right)_{xz}\nabla_{z}\nabla_{z}B_{x}^{*}+\left(Q^{m}\right)_{xx}\nabla_{z}\nabla_{x}B_{x}^{*}+\left(Q^{m}\right)_{xy}\nabla_{z}\nabla_{y}B_{x}^{*}\right],\nonumber \\
 & + & \frac{1}{12}\mathrm{Re}\left[\left(Q^{m}\right)_{zx}\nabla_{z}\nabla_{x}B_{z}^{*}+\left(Q^{m}\right)_{zy}\nabla_{z}\nabla_{y}B_{z}^{*}+\left(Q^{m}\right)_{zz}\nabla_{z}\nabla_{z}B_{z}^{*}\right],\nonumber \\
 & = & \frac{1}{12}\mathrm{Re}\left[\left(Q^{m}\right)_{xz}\nabla_{z}\nabla_{z}B_{x}^{*}+\left(Q^{m}\right)_{zx}\nabla_{z}\nabla_{x}B_{z}^{*}\right]+\frac{1}{12}\mathrm{Re}\left[\left(Q^{m}\right)_{xx}\nabla_{z}\nabla_{z}B_{x}^{*}+\left(Q^{m}\right)_{zz}\nabla_{z}\nabla_{z}B_{z}^{*}\right].\nonumber
\end{eqnarray}

Now, by substituting Eq.~\ref{eq:Derv_B} and Eq.~\ref{eq:Qm} into
Eq.~\ref{eq:FQe_x}, we obtain:

\begin{eqnarray*}
F_{z(Q^{m})} & = & \frac{1}{12}\mathrm{Re}\left\{ \mathrm{i}\alpha_{Q^{m}}\frac{E_{0}}{2}\left[\frac{k^{2}}{\omega\mu}\left(\mathrm{sin^{2}\psi-cos^{2}\psi}\right)\mathrm{cos}\delta\right]\left[E_{0}^{*}\left(\frac{k^{3}}{\omega}\mathrm{cos^{3}\psi cos\delta}\right)\right]\right\} \\
 & + & \frac{1}{12}\mathrm{Re}\left\{ \mathrm{i}\alpha_{Q^{m}}\frac{E_{0}}{2}\left[\frac{k^{2}}{\omega\mu}\left(\mathrm{sin^{2}\psi-cos^{2}\psi}\right)\mathrm{cos}\delta\right]\left[-E_{0}^{*}\left(\frac{k^{3}}{\omega}\mathrm{sin^{2}\psi cos\psi cos\delta}\right)\right]\right\} \\
 & + & \frac{1}{12}\mathrm{Re}\left[\alpha_{Q^{m}}E_{0}\frac{k^{2}}{\omega\mu}\mathrm{sin}\psi\textrm{\ensuremath{\mathrm{cos}\psi}}\mathrm{sin}\delta\left(-\mathrm{i}\right)E_{0}^{*}\left(\frac{k^{3}}{\omega}sin\psi\mathrm{cos^{2}\psi sin\delta}\right)\right]\\
 & + & \frac{1}{12}\mathrm{Re}\left[-\mathrm{i}\alpha_{Q^{m}}E_{0}\frac{k^{2}}{\omega\mu}\mathrm{sin\psi}\textrm{\ensuremath{\mathrm{cos}\psi}}\mathrm{sin}\delta E_{0}^{*}\left(\frac{k^{3}}{\omega}\mathrm{cos^{2}\psi sin\psi sin\delta}\right)\right]\\
 & = & 5F^{\mathrm{norm}}\mathrm{cos\psi\left(\mathrm{cos^{2}2\psi}\mathrm{cos^{2}}\delta+\mathrm{sin^{2}}2\psi\mathrm{sin^{2}}\delta\right)}\mathrm{Im}\left(\frac{k^{5}}{120\pi}\alpha_{Q^{m}}\right)
\end{eqnarray*}

Finally, the normalized contribution derives as:

\begin{equation}
\boxed{\bar{F}_{z(Q^{m})}=5\mathrm{cos\psi\left(\mathrm{cos^{2}2\psi}\mathrm{cos^{2}}\delta+\mathrm{sin^{2}}2\psi\mathrm{sin^{2}}\delta\right)}\mathrm{Im}\left(\bar{\alpha}_{Q^{m}}\right).}
\end{equation}

This expression is documented in Eq.~7 of the main manuscript.

\subsection{$\mathbf{F}_{mQ^{m}}$ contribution }

According to Eq.~\ref{eq:a6}, the term due to the interference of
the magnetic dipole ($m$) and quadrupole($Q^{m}$) is given by

\begin{equation}
F_{i(mQ^{m})}=-\frac{k^{5}}{120\pi\varepsilon_{0}c^{2}}\mathrm{Im}\left[\left(Q^{m}\right)_{ij}m_{j}^{*}\right].\label{eq:11}
\end{equation}

Using Eq.~\ref{eq:11}, the $x$ component of $\mathbf{F}_{mQ^{m}}$
reads as:

\begin{eqnarray}
F_{x(mQ^{m})} & = & -\frac{k^{5}}{120\pi\varepsilon_{0}c^{2}}\mathrm{Im}\left[\left(Q^{m}\right)_{xz}m_{z}^{*}+\left(Q^{m}\right)_{xx}m_{x}^{*}\right],\label{eq:FmQm_x}
\end{eqnarray}

Now, by substituting Eq.~\ref{eq:Qm} and Eq.~\ref{eq:MD} into
Eq.~\ref{eq:FmQm_x}, we obtain:

\begin{eqnarray*}
F_{x(mQ^{m})} & = & -\frac{k^{5}}{120\pi\varepsilon_{0}c^{2}}\mathrm{Im}\left\{ \mathrm{i}\alpha_{Q^{m}}\frac{E_{0}}{2}\left[\frac{k^{2}}{\omega\mu}\left(\mathrm{sin^{2}\psi-cos^{2}\psi}\right)\mathrm{cos}\delta\right]e^{\mathrm{i}k\mathrm{cos}\psi z_{0}}m_{z}^{*}\right\} ,\\
 &  & -\frac{k^{5}}{120\pi\varepsilon_{0}c^{2}}\mathrm{Im}\left(\alpha_{Q^{m}}E_{0}\frac{k^{2}}{\omega\mu}\mathrm{sin}\psi\mathrm{cos}\psi\mathrm{sin}\delta e^{\mathrm{i}k\mathrm{cos}\psi z_{0}}m_{x}^{*}\right),\\
 & = & \frac{3}{2}F^{\mathrm{norm}}\mathrm{sin\psi sin2\delta\left[\mathrm{cos}\left(2\psi\right)+2\mathrm{cos^{2}}\psi\right]}\mathrm{Im}\left(\frac{k^{5}}{120\pi}\frac{k^{3}}{6\pi}\alpha_{Q^{m}}\alpha_{m}^{*}\right).
\end{eqnarray*}

Finally, by using the definition of the normalized force and the normalized
polarizabilities, we obtain:

\begin{equation}
\boxed{\bar{F}_{x(mQ^{m})}=\frac{3}{2}\mathrm{sin\psi sin2\delta\left[\mathrm{cos}\left(2\psi\right)+2\mathrm{cos^{2}}\psi\right]}\mathrm{Im}\left(\bar{\alpha}_{Q^{m}}\bar{\alpha}_{m}^{*}\right).}
\end{equation}

This expression is documented in Eq.~7 of the main manuscript.

Using Eq.~\ref{eq:11}, the $z$ component of $\mathbf{F}_{mQ^{m}}$
read as

\begin{eqnarray}
F_{z(mQ^{m})} & = & -\frac{k^{5}}{120\pi\varepsilon_{0}c^{2}}\mathrm{Im}\left[\left(Q^{m}\right)_{zx}m_{x}^{*}+\left(Q^{m}\right)_{zz}m_{z}^{*}\right].\label{eq:FmQm_z}
\end{eqnarray}

Now, by substituting Eq.~\ref{eq:Qm} and Eq.~\ref{eq:MD} into
Eq.~\ref{eq:FmQm_x}, we obtain:

\begin{eqnarray*}
F_{z(mQ^{m})} & = & -\frac{k^{5}}{120\pi\varepsilon_{0}c^{2}}\mathrm{Im}\left\{ \mathrm{i}\alpha_{Q^{m}}\frac{E_{0}}{2}\left[\frac{k^{2}}{\omega\mu}\left(\mathrm{sin^{2}\psi-cos^{2}\psi}\right)\mathrm{cos}\delta\right]e^{ik\mathrm{cos}\psi z_{0}}m_{x}^{*}\right\} \\
 &  & -\frac{k^{5}}{120\pi\varepsilon_{0}c^{2}}\mathrm{Im}\left(-\alpha_{Q^{m}}E_{0}\frac{k^{2}}{\omega\mu}\mathrm{sin\psi}\mathrm{cos}\psi\mathrm{sin}\delta e^{ik\mathrm{cos}\psi z_{0}}m_{z}^{*}\right),\\
 & = & -3F^{\mathrm{norm}}\mathrm{cos\psi}\left(\mathrm{cos}2\psi\mathrm{cos^{2}}\delta+2\mathrm{sin^{2}}\psi\mathrm{sin^{2}}\delta\right)\mathrm{Re}\left(\frac{k^{5}}{120\pi}\frac{k^{3}}{6\pi}\alpha_{Q^{m}}\alpha_{m}^{*}\right).
\end{eqnarray*}

Finally, the normalized contribution is derived as:

\begin{equation}
\boxed{\bar{F}_{z(mQ^{m})}=-3\mathrm{cos\psi}\left(\mathrm{cos}2\psi\mathrm{cos^{2}}\delta+2\mathrm{sin^{2}}\psi\mathrm{sin^{2}}\delta\right)\mathrm{Re}\left(\bar{\alpha}_{Q^{m}}\bar{\alpha}_{m}^{*}\right).}
\end{equation}

This expression is documented in Eq.~7 of the main manuscript.

\subsection{$\mathbf{F}_{Q^{e}Q^{m}}$ contribution }

According to Eq.~\ref{eq:a6}, the term due to the interference of
the electric quadrupole ($Q^{e}$) and and magnetic quadrupole ($Q^{m}$)
is given by:

\begin{equation}
F_{i(Q^{e}Q^{m})}=-\frac{k^{6}}{9\times240\pi\varepsilon_{0}c}\mathrm{Re}\left[\underset{l,j,k}{\sum}\varepsilon_{ijk}(Q^{e})_{lj}\left(Q^{m}\right)_{lk}^{*}\right].\label{eq:12}
\end{equation}

Using Eq.~\ref{eq:11}, the $x$ component of $\mathbf{F}_{Q^{e}Q^{m}}$
reads as:

\begin{eqnarray}
F_{x(Q^{e}Q^{m})} & = & -\frac{k^{6}}{9\times240\pi\varepsilon_{0}c}\mathrm{Re}\left[\underset{l,j,k}{\sum}\varepsilon_{1jk}(Q^{e})_{lj}\left(Q^{m}\right)_{lk}^{*}\right],\nonumber \\
 & = & -\frac{k^{6}}{9\times240\pi\varepsilon_{0}c}\mathrm{Re}\left\{ \left[\underset{l}{\sum}\varepsilon_{123}(Q^{e})_{l2}\left(Q^{m}\right)_{l3}^{*}\right]+\left[\underset{l}{\sum}\varepsilon_{132}(Q^{e})_{l3}\left(Q^{m}\right)_{l2}^{*}\right]\right\} ,\nonumber \\
 & = & -\frac{k^{6}}{9\times240\pi\varepsilon_{0}c}\mathrm{Re}\left\{ (+1).\left[Q_{12}^{e}Q_{13}^{m*}+Q_{22}^{e}Q_{23}^{m*}+Q_{32}^{e}Q_{33}^{m*}\right]\right.\nonumber \\
 &  & \left.+(-1)\left[Q_{13}^{e}Q_{12}^{m*}+Q_{23}^{e}Q_{22}^{m*}+Q_{33}^{e}Q_{32}^{m*}\right]\right\} ,\nonumber \\
 & = & -\frac{k^{6}}{9\times240\pi\varepsilon_{0}c}\mathrm{Re}\left(Q_{xy}^{e}Q_{xz}^{m*}+Q_{zy}^{e}Q_{zz}^{m*}\right).\label{eq:FQeQm_x}
\end{eqnarray}

Now, by substituting Eq.~\ref{eq:Qe} and Eq.~\ref{eq:Qm} into
Eq.~\ref{eq:FQeQm_x}, we obtain:

\begin{eqnarray*}
F_{x(Q^{e}Q^{m})} & = & -\frac{5}{6}F^{\mathrm{norm}}\mathrm{sin\psi\mathrm{sin}2\delta}\left(\mathrm{cos}2\psi+2\mathrm{cos^{2}}\psi\right)\mathrm{Im}\left[\left(\frac{k^{5}}{120\pi}\right)^{2}\alpha_{Q^{e}}\alpha_{Q^{m}}^{*}\right].
\end{eqnarray*}

Finally, the normalized contribution is derived as:

\[
\boxed{\bar{F}_{x(Q^{e}Q^{m})}=-\frac{5}{6}\mathrm{sin\psi\mathrm{sin}2\delta}\left(\mathrm{cos}2\psi+2\mathrm{cos^{2}}\psi\right)\mathrm{Im}\left(\bar{\alpha}_{Q^{e}}\bar{\alpha}_{Q^{m}}^{*}\right).}
\]

This expression is documented in Eq.~7 of the main manuscript.

Using Eq.~\ref{eq:11}, the $z$ component of $\mathbf{F}_{Q^{e}Q^{m}}$
read as

\begin{eqnarray}
F_{z(Q^{e}Q^{m})} & = & -\frac{k^{6}}{9\times240\pi\varepsilon_{0}c}\mathrm{Re}\left[\underset{l,j,k}{\sum}\varepsilon_{3jk}(Q^{e})_{lj}\left(Q^{m}\right)_{lk}^{*}\right]\nonumber \\
 & = & -\frac{k^{6}}{9\times240\pi\varepsilon_{0}c}\mathrm{Re}\left[\underset{l}{\sum}\varepsilon_{312}(Q^{e})_{l1}\left(Q^{m}\right)_{l2}^{*}+\underset{l}{\sum}\varepsilon_{321}(Q^{e})_{l2}\left(Q^{m}\right)_{l1}^{*}\right]\nonumber \\
 & = & -\frac{k^{6}}{9\times240\pi\varepsilon_{0}c}\mathrm{Re}\left\{ \left(+1\right)\left[Q_{11}^{e}Q_{12}^{m*}+Q_{21}^{e}Q_{22}^{m*}+Q_{31}^{e}Q_{32}^{m*}\right]\right.\nonumber \\
 &  & \left.+(-1)\left[Q_{12}^{e}Q_{11}^{m*}+Q_{22}^{e}Q_{21}^{m*}+Q_{32}^{e}Q_{31}^{m*}\right]\right\} ,\nonumber \\
 & = & \frac{k^{6}}{9\times240\pi\varepsilon_{0}c}\mathrm{Re}\left(Q_{zy}^{e}Q_{zx}^{m*}+Q_{xy}^{e}Q_{xx}^{m*}\right).\label{eq:FQeQm_z}
\end{eqnarray}

Now, by substituting Eq.~\ref{eq:Qe} and Eq.~\ref{eq:Qm} into
Eq.~\ref{eq:FQeQm_z}, we obtain

\begin{eqnarray*}
F_{z(Q^{e}Q^{m})} & = & \frac{k^{6}}{9\times240\pi\varepsilon_{0}c}\mathrm{Re}\left(Q_{zy}^{e}Q_{zx}^{m*}+Q_{xy}^{e}Q_{xx}^{m*}\right)\\
 & =- & \frac{5}{3}F^{\mathrm{norm}}\mathrm{cos\psi\left(\mathrm{cos}2\psi\mathrm{cos^{2}}\delta+2\mathrm{sin^{2}\psi}\mathrm{sin^{2}}\delta\right)}\mathrm{Re}\left[\left(\frac{k^{5}}{120\pi}\right)^{2}\alpha_{Q^{e}}\alpha_{Q^{m}}^{*}\right].
\end{eqnarray*}

Finally, the normalized contribution is derived as:

\begin{equation}
\boxed{\bar{F}_{z(Q^{e}Q^{m})}=-\frac{5}{3}\mathrm{cos\psi}\left(\mathrm{cos}2\psi\mathrm{cos^{2}}\delta+2\mathrm{sin^{2}\psi}\mathrm{sin^{2}}\delta\right)\mathrm{Re}\left(\bar{\alpha}_{Q^{e}}\bar{\alpha}_{Q^{m}}^{*}\right).}
\end{equation}

This expression is documented in Eq.~7 of the main manuscript.

\section{Optical force: TM illumination}

Using the duality in the Maxwell\textquoteright s equations for the
electric and magnetic fields/induced moments, similar expression for
optical force can be obtained for a TM polarization. The results are
as following:

\begin{eqnarray}
\overline{\mathbf{F}}^{\mathrm{TM}} & \approx & \mathbf{\overline{F}}_{p}^{\mathrm{TM}}+\mathbf{\overline{F}}_{m}^{\mathrm{TM}}+\mathbf{\overline{F}}_{pm}^{\mathrm{TM}}+\mathbf{\overline{F}}_{Q^{e}}^{\mathrm{TM}}+\mathbf{\overline{F}}_{Q^{m}}^{\mathrm{TM}}+\mathbf{\overline{F}}_{pQ^{e}}^{\mathrm{TM}}+\mathbf{\overline{F}}_{mQ^{m}}^{\mathrm{TM}}+\mathbf{\overline{F}}_{Q^{e}Q^{m}}^{\mathrm{TM}},\label{eq:Force_TM}\\
\mathbf{\overline{F}}_{p}^{\mathrm{TM}} & = & -\frac{3}{2}\mathrm{sin\psi}\mathrm{sin}2\delta\mathrm{cos2}\psi\mathrm{Re}\left(\bar{\alpha_{e}}\right)\mathbf{e}_{x}+3\mathrm{cos}\psi\mathrm{\left(\mathrm{cos^{2}\psi cos^{2}}\delta+\mathrm{sin^{2}}\psi\mathrm{sin^{2}}\delta\right)\mathrm{Im}\left(\bar{\alpha_{e}}\right)}\mathbf{e}_{z},\nonumber \\
\mathbf{\overline{F}}_{m}^{\mathrm{TM}} & = & -\frac{3}{2}\mathrm{sin\psi}\mathrm{sin}2\delta\mathrm{Re}\left(\bar{\alpha_{m}}\right)\mathbf{e}_{x}+3\mathrm{cos}\psi\mathrm{cos^{2}}\delta\mathrm{Im}\left(\bar{\alpha_{m}}\right)\mathbf{e}_{z},\nonumber \\
\mathbf{\overline{F}}_{pm}^{\mathrm{TM}} & = & +\frac{3}{2}\mathrm{sin\psi}\mathrm{sin}2\delta\mathrm{Im}\left(\bar{\alpha_{e}}\bar{\alpha}_{m}^{*}\right)\mathbf{e}_{x}-3\mathrm{cos}\psi\mathrm{cos^{2}}\delta\mathrm{Re}\left(\bar{\alpha_{e}}\bar{\alpha}_{m}^{*}\right)\mathbf{e}_{z},\nonumber \\
\mathbf{\overline{F}}_{Q^{e}}^{\mathrm{TM}} & = & -\frac{5}{2}\mathrm{sin}\psi\mathrm{sin}2\delta\mathrm{cos}4\psi\mathrm{Re}\left(\bar{\alpha}_{Q^{e}}\right)\mathbf{e}_{x}+5\mathrm{cos}\psi\left(\mathrm{cos^{2}2\psi cos^{2}}\delta+\mathrm{sin^{2}2\psi sin^{2}}\delta\right)\mathrm{Im}\left(\bar{\alpha}_{Q^{e}}\right)\mathbf{e}_{z},\nonumber \\
\mathbf{\overline{F}}_{Q^{m}}^{\mathrm{TM}} & = & -\frac{5}{2}\mathrm{sin}\psi\mathrm{sin}2\delta\mathrm{cos2\psi}\mathrm{Re}\left(\bar{\alpha}_{Q^{m}}\right)\mathbf{e}_{x}+5\mathrm{cos}\psi\left(\mathrm{cos^{2}\psi cos^{2}}\delta+\mathrm{sin^{2}}\psi\mathrm{sin^{2}}\delta\right)\mathrm{Im}\left(\bar{\alpha}_{Q^{m}}\right)\mathbf{e}_{z},\nonumber \\
\mathbf{\overline{F}}_{pQ^{e}}^{\mathrm{TM}} & = & +\frac{3}{2}\mathrm{sin}\psi\mathrm{sin}2\delta\left(\mathrm{cos2\psi+2cos^{2}\psi}\right)\mathrm{Im}\left(\bar{\alpha}_{Q^{e}}\bar{\alpha}_{e}^{*}\right)\mathbf{e}_{x}-3\mathrm{cos}\psi\left(\mathrm{cos^{2}}\delta\mathrm{cos}2\psi+2\mathrm{sin^{2}}\psi\mathrm{sin^{2}}\delta\right)\mathrm{Re}\left(\bar{\alpha}_{Q^{e}}\bar{\alpha}_{e}^{*}\right)\mathbf{e}_{z},\nonumber \\
\mathbf{\overline{F}}_{mQ^{m}}^{\mathrm{TM}} & = & +\frac{3}{2}\mathrm{sin}\psi\mathrm{sin}2\delta\mathrm{Im}\left(\bar{\alpha}_{Q^{m}}\bar{\alpha}_{m}^{*}\right)\mathbf{e}_{x}-3\mathrm{cos}\psi\mathrm{cos^{2}}\delta\mathrm{Re}\left(\bar{\alpha}_{Q^{m}}\bar{\alpha}_{m}^{*}\right)\mathbf{e}_{z},\nonumber \\
\overline{\mathbf{F}}_{Q^{e}Q^{m}}^{\mathrm{TM}} & = & +\frac{5}{6}\mathrm{sin}\psi\mathrm{sin}2\delta\left(\mathrm{\mathrm{cos2\psi+2cos^{2}\psi}}\right)\mathrm{Im}\left(\bar{\alpha}_{Q^{e}}\bar{\alpha}_{Q^{m}}^{*}\right)\mathbf{e}_{x}-\frac{5}{3}\mathrm{cos}\psi\left(\mathrm{cos^{2}}\delta\mathrm{cos}2\psi+2\mathrm{sin^{2}\psi sin^{2}}\delta\right)\mathrm{Re}\left(\bar{\alpha}_{Q^{e}}\bar{\alpha}_{Q^{m}}^{*}\right)\mathbf{e}_{z}.\nonumber
\end{eqnarray}

\end{widetext}


%

\end{document}